\documentclass[journal]{IEEEtran}
\usepackage{amsmath,amsfonts}
\usepackage{algorithmic}
\usepackage{algorithm}
\usepackage{array}
\usepackage[caption=false,font=normalsize,labelfont=sf,textfont=sf]{subfig}
\usepackage{textcomp}
\usepackage{stfloats}
\usepackage{url}
\usepackage{verbatim}
\usepackage{graphicx}
\usepackage{cite}
\usepackage{enumitem}
\usepackage{bm}
\usepackage{amsthm} 
\usepackage{CJK}
\usepackage{indentfirst}
\usepackage{cases}
\usepackage{tabularx}
\usepackage{xurl} 
\theoremstyle{definition}
\newtheorem{definition}{Definition}

\begin{document}

\title{Spatial Signal Focusing and Noise Suppression for Direction-of-Arrival Estimation in Large-Aperture 2D Arrays under Demanding Conditions}

\author{Xuyao Deng, Yong Dou, Kele Xu,~\IEEEmembership{Senior Member,~IEEE} 
}



\maketitle

\begin{abstract}
Direction-of-Arrival (DOA) estimation in sensor arrays faces limitations under demanding conditions, including low signal-to-noise ratio, single-snapshot scenarios, coherent sources, and unknown source counts. Conventional beamforming suffers from sidelobe interference, adaptive methods (e.g., MVDR) and subspace algorithms (e.g., MUSIC) degrade with limited snapshots or coherent signals, while sparse-recovery approaches (e.g., L1-SVD) incur high computational complexity for large arrays. In this article, we construct the concept of the optimal spatial filter to solve the DOA estimation problem under demanding conditions by utilizing the sparsity of spatial signals. By utilizing the concept of the optimal spatial filter, we have transformed the DOA estimation problem into a solution problem for the optimal spatial filter. We propose the Spatial Signal Focusing and Noise Suppression (SSFNS) algorithm, which is a novel DOA estimation framework grounded in the theoretical existence of an optimal spatial filter, to solve for the optimal spatial filter and obtain DOA. Through experiments, it was found that the proposed algorithm is suitable for large aperture two-dimensional arrays and experiments have shown that our proposed algorithm performs better than other algorithms in scenarios with few snapshots or even a single snapshot, low signal-to-noise ratio, coherent signals, and unknown signal numbers in two-dimensional large aperture arrays.

\end{abstract}


\section{Introduction}

\IEEEPARstart{E}{stimating} Direction of Arrival (DOA) under demanding conditions including low signal-to-noise ratio, single-snapshot scenarios, coherent sources, and unknown source counts has important applications in extracting information conveyed by propagating waves in fields such as radar, sonar, wireless communications, and geophysics~\cite{krishnaveni2013beamforming,chung2014doa,zheng2024deep,kulkarni2025comprehensive}. Consequently, DOA estimation under demanding conditions, through processing data measured by sensor arrays, has attracted significant attention from numerous researchers over the past several decades~\cite{fortunati2014single,shen2016underdetermined,bilgehan2019fast,papageorgiou2021deep,ge2021deep,wei2022fast,molaei2024comprehensive,muqaibel2024sparse,kulkarni2025comprehensive}.

One of the classic methods for estimating DOA is beamforming technology, which processes signals in conjunction with sensor array measurements in the presence of interference and noise. Functioning as spatial filters~\cite{van1988beamforming}, beamforming methods are broadly categorized into conventional and adaptive beamforming. In conventional beamforming (CBF)~\cite{krim1996two}, they are data-independent beamformers, exhibiting a constant response across all signal and interference scenarios. Unlike conventional beamforming, adaptive beamformers are data-dependent, as their weight vectors are computed based on the incoming data to optimize performance under various constraints~\cite{harry2002detection,chung2014doa}. A well-known representative method is the Minimum Variance Distortionless Response (MVDR) beamformer~\cite{capon2005high}. However, when performing spatial weighted filtering, CBF introduces undesirable contributions from signals or interference arriving from other directions due to the presence of sidelobes. This effect is particularly pronounced when the number of array sensor elements is small, as mutual interference between signals from different directions intensifies, potentially causing the effectiveness of CBF to be significantly impaired. Furthermore, traditional adaptive beamforming algorithms suffer from degradation if certain assumptions about the environment, signal sources, or the sensor array become invalid or inaccurate~\cite{krishnaveni2013beamforming}. Similar degradation occurs when the available sample size is insufficient. For instance, in the case of the MVDR method, performance degrades when either of the following conditions exists: (1) the number of samples used for computing the sample covariance matrix is less than the number of array sensor elements, or (2) coherent signal sources are encountered. This degradation stems from rank deficiency problems encountered when inverting the received signal covariance matrix.

\IEEEpubidadjcol
Subspace method is a type of method in DOA estimation. The Multiple Signal Classification (MUSIC)~\cite{bienvenu1983optimality,schmidt1986multiple} algorithm is a prominent representative of subspace methods. MUSIC involves performing eigenvalue decomposition (EVD) on the covariance matrix of the array output data. This decomposition yields the signal subspace corresponding to the signal components, and the noise subspace, which is the orthogonal complement to the signal subspace. The algorithm then utilizes the properties of these two subspaces to estimate signal parameters. The introduction of MUSIC spurred the emergence and development of eigenstructure-based methods. To reduce complexity and enhance performance and resolution, several MUSIC variants were subsequently proposed, including Unitary MUSIC~\cite{pesavento2002unitary} and Root MUSIC~\cite{rao1989performance}. However, the performance of subspace methods degrades significantly in the presence of correlated source signals. This degradation occurs because the signal subspace suffers from rank deficiency under such conditions. Additionally, subspace algorithms require prior knowledge of the number of signal sources. Mistakes in source number estimation – either overestimation or underestimation – can lead to the MUSIC spatial spectrum exhibiting spurious peaks or the vanishing of true spectral peaks~\cite{zhang2009music}.

Furthermore, in the single-snapshot scenario, representative adaptive beamforming methods like MVDR and subspace methods like MUSIC become ineffective. This failure occurs because the limited snapshot data prevents the accurate estimation of the covariance matrix and also leads to rank deficiency in the estimated covariance matrix.

Another type method for estimating DOA is based on sparse representation. With the advancement of sparse recovery theory and methodologies~\cite{candes2006near,candes2006robust,donoho2006compressed,chen2001atomic}, sparse representation-based methods have garnered significant attention from researchers~\cite{tropp2006just}. Given that the number of signals in array processing is often limited, the methods proposed in~\cite{fuchs2001application,malioutov2005sparse} reformulate DOA estimation as a sparse recovery problem, assigning non-zero amplitudes to signal components at specific angles. A representative method, L1-SVD~\cite{malioutov2005sparse}, leverages the sparsity of signal sources in the spatial domain to construct a sparse representation. It then frames the problem as an optimization framework (specifically related to the Lasso problem~\cite{tibshirani1996regression}) to identify the non-zero components. The DOA estimates are subsequently derived from the angles associated with these identified non-zero components. However, methods based on sparse representations generally exhibit higher computational complexity compared to beamforming and subspace methods. Particularly as the number of array sensor elements increases, the algorithmic computational load rises substantially, making it difficult to guarantee real-time performance.

In order to address the inherent limitations of existing methods—namely, to eliminate the adverse effects of signals or interference from other directions introduced by side lobes in conventional beamformers; to overcome the failure of traditional adaptive beamforming approaches (like MVDR) and subspace-based methods (such as MUSIC) under conditions of a single or limited number of snapshots and the presence of coherent sources; to avoid the high computational complexity associated with sparsity-based representation methods (e.g., L1-SVD), which results in poor real-time DOA estimation performance for arrays with a large number of sensor elements; and to circumvent the requirement inherent in subspace methods like MUSIC for an accurate estimate of the source number—we propose a novel algorithm named the Spatial Signal Focusing and Noise Suppression algorithm. This enables DOA estimation with low computational complexity under demanding conditions—including without requiring knowledge of the source number, with a low or a single snapshot count, with coherent sources present, and in the presence of interfering signals from other directions.

Through analysis (see Section~\ref{subsec:Optimal Spatial Filter}), we find that, under the single-snapshot condition and assuming the complete array manifold matrix and the noise are linearly independent, there exists a spatial filter that can completely eliminate noise while simultaneously separating signals arriving from different directions in space. We refer to this as the optimal spatial filter. However, due to the unknown nature of noise, this optimal spatial filter is difficult to obtain directly. The goal of our proposed algorithm is to strive to achieve the optimal spatial filter or a close approximation to it. We formulate the solution for the optimal spatial filter as an optimization problem and solve it by iteratively introducing different constraints. The approach proceeds by first obtaining an initial spatial filter that suppresses noise and simultaneously separates signals from different spatial directions as much as possible. Then, it iteratively refines this filter by introducing constraints based on angles associated with the maximum values in the resulting spatial power spectrum after filtering. Our final goal is to obtain or approximate the optimal spatial filter that can eliminate noise while separating signals from different directions in space (detailed in Section~\ref{sec:Spatial Signal Focusing and Noise Suppression Algorithm}).

The Spatial Signal Focusing and Noise Suppression algorithm ultimately yields a spatial filter capable of achieving DOA estimation under single-snapshot conditions. Significantly, the proposed algorithm is unaffected by coherent signal sources and can perform DOA estimation without prior knowledge of the number of signal sources. When compared using the large aperture 2D array configuration, our method delivers superior DOA estimation performance at low Signal-to-Noise Ratios (SNR) compared to representative algorithms including MVDR, MUSIC, and L1-SVD.

The primary contributions of this paper are summarized as follows:
\begin{enumerate}[label=\alph*)]
    \item \textbf{Theoretical Foundation}: Through rigorous mathematical analysis, we establish that under the single-snapshot condition and assuming linear independence between the complete array manifold matrix and noise, there exists a spatial filter capable of simultaneously eliminating noise and separating signals impinging from distinct directions in space. We term this filter the optimal spatial filter. This analysis provides the theoretical underpinning for utilizing a variety of techniques to solve for such spatial filters.

    \item \textbf{Proposed Algorithm}: We introduce the Spatial Signal Focusing and Noise Suppression algorithm to solve for this spatial filter. The resulting filter achieves DOA estimation under single-snapshot conditions. Crucially, the algorithm operates effectively even in the presence of coherent signal sources and without requiring prior knowledge of the number of sources. Furthermore, for arrays with a large number of sensor elements, the proposed algorithm achieves DOA estimation with low computational complexity and demonstrates better performance at low Signal-to-Noise Ratios (SNR).

    \item \textbf{Performance Evaluation Metrics}: Due to the unknown nature of noise, the solution typically yields an approximation of the optimal spatial filter. To quantify the deviation between the obtained filter and the optimal one, we introduce novel metrics specifically designed to evaluate spatial filter performance. Furthermore, for large-aperture arrays where high angular resolution is inherent, we propose novel performance evaluation metrics that jointly assess both angular resolution capability and DOA estimation accuracy.

    \item \textbf{Open-source code}: We have open-source our proposed algorithm code on GitHub to facilitate research~\footnote{\url{https://github.com/colaudiolab/Spatial-Signal-Focusing-and-Noise-Suppression-Algorithm}}.
\end{enumerate}

\section{System Model And Preliminaries}
In this section, the system model will be described in detail to provide the signal model and optimization objective for introducing the Spatial Signal Focusing and Noise Suppression algorithm in subsequent sections. First, we discuss the signal model. Then, the mathematical formulation of the DOA estimation problem is presented. Subsequently, the fundamental framework for solving the problem via the Spatial Signal Focusing and Noise Suppression algorithm is established. Finally, the definition of the optimal spatial filter is provided.

\subsection{Signal Model}
\label{sec:Signal_Model}
Consider $K$ far-field narrowband signals impinging on a spatial array. The array comprises $M$ sensor elements. Here, it is assumed that the number of sensor elements equals the number of channels. That is, after receiving the signals, each sensor element transmits its signal to the processor through its respective transmission channel. Thus, the processor receives data from $M$ channels. Under the narrowband signal source assumption, the $i$-th signal can be expressed in complex envelope form as follows:
\begin{equation}
\label{eq:complex_envelope_form}
\begin{cases}
s_i(t)=u_i(t)\mathrm{e}^{\mathrm{j}(\omega_0t+\varphi(t))} \\
s_i(t-\tau)=u_i(t-\tau)\mathrm{e}^{\mathrm{j}(\omega_0(t-\tau)+\varphi(t-\tau))} &.
\end{cases}
\end{equation}

In the Formula~\eqref{eq:complex_envelope_form}, $u_i(t)$ is the amplitude of the received signal, $\varphi(t)$ is the phase of the received signal, $\tau$ is the time delay, and $\omega_0$ is the angular frequency of the received signal.

Under the assumption of narrowband, far-field signal sources, we have:
\begin{equation}
\label{eq:u_i(t)&phi(t)_approximation}
\begin{cases}
u_i(t-\tau)\approx u_i(t) \\
\varphi(t-\tau)\approx\varphi(t) &.
\end{cases}
\end{equation}

According to Formulas~\eqref{eq:complex_envelope_form} and~\eqref{eq:u_i(t)&phi(t)_approximation}, the following formula holds:
\begin{equation}
s_i(t-\tau)\approx s_i(t)\mathrm{e}^{-j\omega_0\tau}\quad i=1,2,\cdots,K.
\end{equation}

Then the received signal of the $j$-th sensor element can be obtained as:
\begin{equation}
x_j(t)=\sum_{i=1}^Kg_{ji}s_i(t-\tau_{ji})+n_j(t)\quad j=1,2,\cdots,M,
\end{equation}
where $g_{ji}$ is the gain coefficient of the $j$-th sensor element for the $i$-th signal, $n_j(t)$ represents the additive noise component at the $j$-th sensor element at time $t$, and $\tau_{ji}$ denotes the time delay of the $i$-th signal arriving at the $j$-th sensor element relative to the reference sensor element.

By arranging the signals received by $M$ array elements at time $t$ into a column vector, we can obtain:
\begin{equation}
\label{eq:x=gs+n}
\begin{bmatrix}
x_1(t) \\
x_2(t) \\
\vdots \\
x_M(t)
\end{bmatrix}=
\begin{bmatrix}
\bm{g}_{1}(\omega_0)...\bm{g}_{K}(\omega_0)
\end{bmatrix}
\begin{bmatrix}
s_1(t) \\
s_2(t) \\
\vdots \\
s_K(t)
\end{bmatrix}+
\begin{bmatrix}
n_1(t) \\
 \\
n_2(t) \\
\vdots \\
n_M(t)
\end{bmatrix}
\end{equation}
of which
\begin{equation}
\label{eq:bf(g)_i(w_0)}
\bm{g}_i(\omega_0)=
\begin{bmatrix}
g_{1i}e^{-j\omega_0\tau_{1i}} \\
g_{2i}e^{-j\omega_0\tau_{2i}} \\
\vdots \\
g_{Mi}e^{-j\omega_0\tau_{Mi}}
\end{bmatrix}\quad i=1,2,...,K.
\end{equation}

Under ideal conditions, assuming all sensor elements in the array are isotropic and unaffected by factors such as channel mismatch or mutual coupling, the gain $g_{ji}$ in Formula~\eqref{eq:bf(g)_i(w_0)} can be omitted (i.e., normalized to 1). With this assumption, Formula~\eqref{eq:x=gs+n} can be simplified as:
\begin{equation}
\label{eq:x=as+n}
\begin{bmatrix}
x_1(t) \\
x_2(t) \\
\vdots \\
x_M(t)
\end{bmatrix}=
\begin{bmatrix}
\bm{a}_{1}(\omega_0)...\bm{a}_{K}(\omega_0)
\end{bmatrix}
\begin{bmatrix}
s_1(t) \\
s_2(t) \\
\vdots \\
s_K(t)
\end{bmatrix}+
\begin{bmatrix}
n_1(t) \\
 \\
n_2(t) \\
\vdots \\
n_M(t)
\end{bmatrix}
\end{equation}
of which
\begin{equation}
\label{eq:bf(a)_i(w_0)}
\bm{a}_i(\omega_0)=
\begin{bmatrix}
e^{-j\omega_0\tau_{1i}} \\
e^{-j\omega_0\tau_{2i}} \\
\vdots \\
e^{-j\omega_0\tau_{Mi}}
\end{bmatrix}\quad i=1,2,...,K
\end{equation}
and
\begin{equation}
    \label{eq:w0=2pif}
    \omega_0 = 2\pi f = \frac{2\pi c}{\lambda},
\end{equation}
where $f$ is frequency, $c$ is speed, and $\lambda$ is wavelength.

Write Formula~\ref{eq:x=as+n} in vector form as follows:
\begin{equation}
\bm{X}(t)=\bm{A}\bm{S}(t)+\bm{N}(t),
\end{equation}
where $\bm{X}(t)$ is the snapshot data vector of the array, $\bm{N}(t)$ is the noise data vector of the array, $\bm{S}(t)$ is the spatial signal vector, and $\bm{A}$ is the array manifold matrix (steering vector matrix):
\begin{equation}
\label{eq:A=[a1,a2,..,ak]}
\bm{A}=
\begin{bmatrix}
\bm{a}_1(\omega_0) & \bm{a}_2(\omega_0) & \cdots & \bm{a}_K(\omega_0).
\end{bmatrix}
\end{equation}

\begin{figure}[t] 
	\centering
	\includegraphics[width=2.8in]{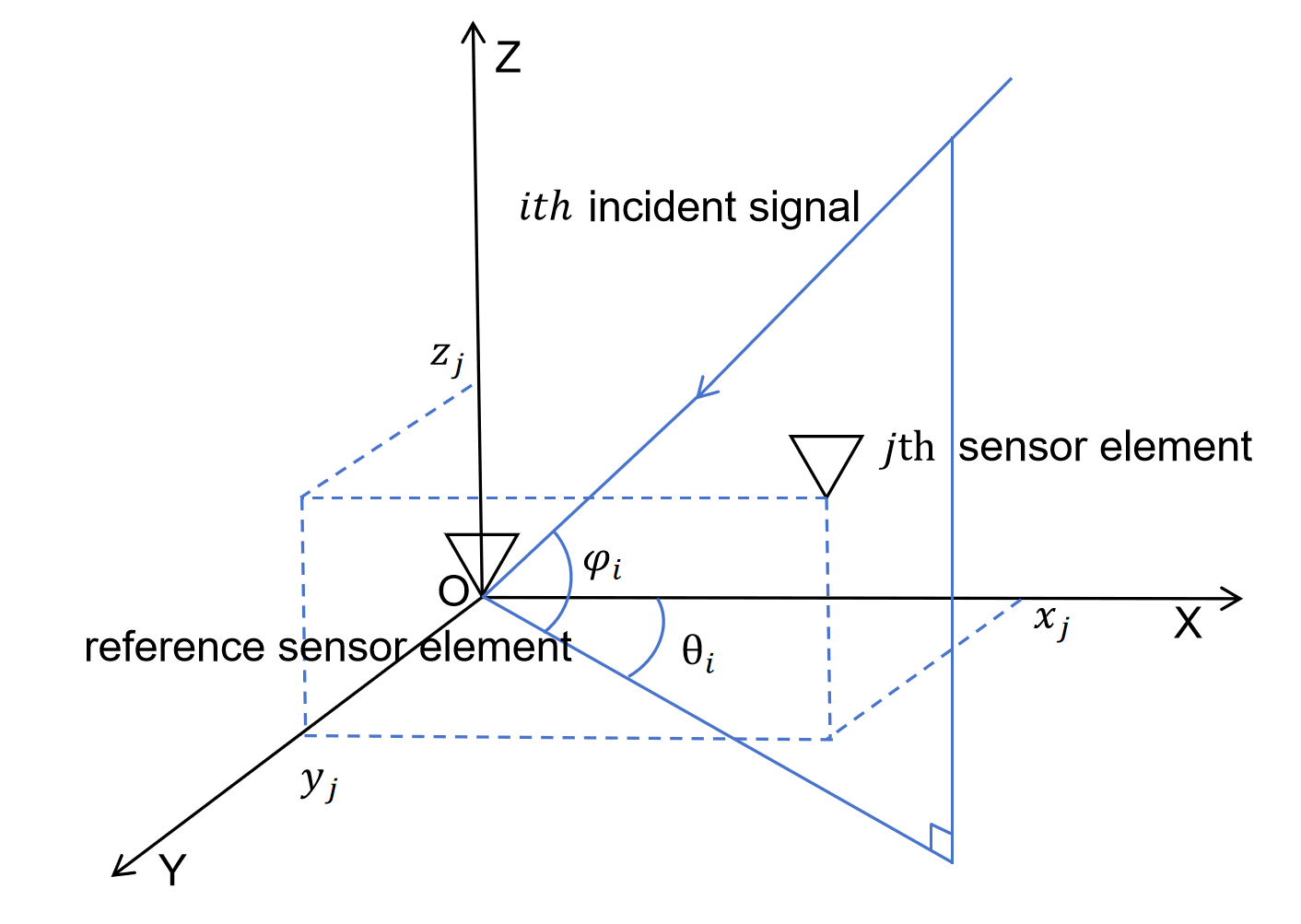}
	\caption{The geometric relationship between the incident signal, the sensor element, and the reference sensor element in space.}
	\label{fig:geometric_relationship}
\end{figure}

Given the time delay $\tau_{ji}$ of each array sensor element relative to a reference sensor element, the array manifold matrix $\bm{A}$ can be computed using Formulas~\eqref{eq:bf(a)_i(w_0)},~\eqref{eq:w0=2pif} and~\eqref{eq:A=[a1,a2,..,ak]}. The expression for the time delays of the array sensor elements in space relative to the reference sensor element is given by:
\begin{equation}
    \label{eq:tau_ji_in_3D}
    \tau_{ji}= \frac{1}{c} \left( x_j \cos \theta_i \cos \varphi_i+y_j \sin \theta_i \cos \varphi_i+z_j \sin \varphi_i \right),
\end{equation}
where $x_j$, $y_j$, $z_j$ are the coordinates of the $j$-th sensor element in the spatial coordinate system with the reference sensor element as the origin; $\theta_i$ and $\varphi_i$ denote the azimuth angle and elevation angle of incidence for the $i$-th signal respectively. The geometric relationship between the incident signal, the sensor element, and the reference sensor element in space is illustrated in Fig. \ref{fig:geometric_relationship}.

For further details, we refer the reader to references~\cite{Wang2004Spatial, tuncer2009narrowband,friedlander2009wireless,chen2010introduction}.

\subsection{Problem Formulation}
DOA estimation utilizes measurements from an array aperture to determine the incident angles of signal sources, thereby enabling signal source localization. Formally, DOA estimation involves estimating signal angles $\theta_s = \{\theta_i~|~\theta_i~is~azimuth~angle~of~ith~signal,i=1\cdots K \}$ and $\varphi_s = \{\varphi_i~|~\varphi_i~is~elevation~angle~of~ith~signal,i=1\cdots K \}$ from the array snapshot data vectors $\bm{X}(t) = [x_1(t), ..., x_M(t)]^T$ measured at multiple time instants $t \in {1, ..., T}$, where T denotes the number of snapshots. The array snapshot vector $\bm{X}(t)$ is obtained from the signal model described in section \ref{sec:Signal_Model} and summarized by Formula~\eqref{eq:x=as+n}.

$\theta_s$ and $\varphi_s$ are the parameters to be estimated in the DOA estimation problem. In this paper, we consider a two-dimensional array configuration. Furthermore, to estimate $\theta_s$ and $\varphi_s$, we can iterate over each possible value of $\varphi$ and solve for the corresponding $\theta_s^{'},~where~\theta_s^{'} \subseteq \theta_s$, under that fixed $\varphi$~\cite{krim2002two}. For each elevation angle $\varphi$ encountered during this iteration, solving for the azimuth angles $\theta_s^{'}$ becomes an identical subproblem. Therefore, we approach the DOA estimation problem by first determining $\varphi$, thereby reducing it to the problem of estimating the azimuth angles $\theta_s^{'}$ with a two-dimensional array given a fixed $\varphi$. To simplify the presentation, we assume $\varphi = 0$ and $\theta_s^{'} = \theta_s$. In the case of a two-dimensional array and $\varphi=0$, Formula~\eqref{eq:tau_ji_in_3D} can be written as:
\begin{equation}
    \label{eq:tau_ji_in_2D}
    \tau_{ji}= \frac{1}{c} \left( x_j \cos \theta_i +y_j \sin \theta_i  \right).
\end{equation}

For a linear array, set $x_j = 0$, we obtain:
\begin{equation}
    \label{eq:tau_ji_in_1D}
    \tau_{ji}= \frac{1}{c} \left( y_j \sin \theta_i  \right).
\end{equation}

We assume that the signals are far-field and narrowband. Furthermore, under ideal conditions, we assume that the array elements are isotropic and not subject to influences such as channel mismatch or mutual coupling. Regarding Formula~\eqref{eq:x=as+n}, we have the received signal $\bm{X}(t)$ at $T$ snapshot time points (comprising $M * T$ data points in total), the number of array elements $M$, and the positions $(x_j,y_j),j=1,...,M$ of the sensor elements. The number of signal sources $K$, the spatial signal vector $\bm{S}(t)$, and the noise data vector $\bm{N}(t)$ in Formula~\eqref{eq:x=as+n} are unknown. The propagation speed $c$ and center frequency $\omega_0$ are empirically determined. 

Therefore, for the DOA estimation problem, it can be summarized as follows: 

\textit{Under the assumptions of a far-field, narrowband signal, ideal array conditions, and $\varphi = 0$ with $\theta_s^{'} = \theta_s$, given the known positional coordinates of $M$ array elements in a two-dimensional array and the array received signal $\bm{X}(t)$ comprising $T$ snapshots, and with propagation speed $c$ and center frequency $\omega_0$ obtained empirically, how to estimate $\theta_s = \{\theta_1, \theta_2, ..., \theta_K\}$, which represents the set of azimuth angles of arrival for the signal sources?}

\begin{figure}[t] 
	\centering
	\includegraphics[width=3.5in]{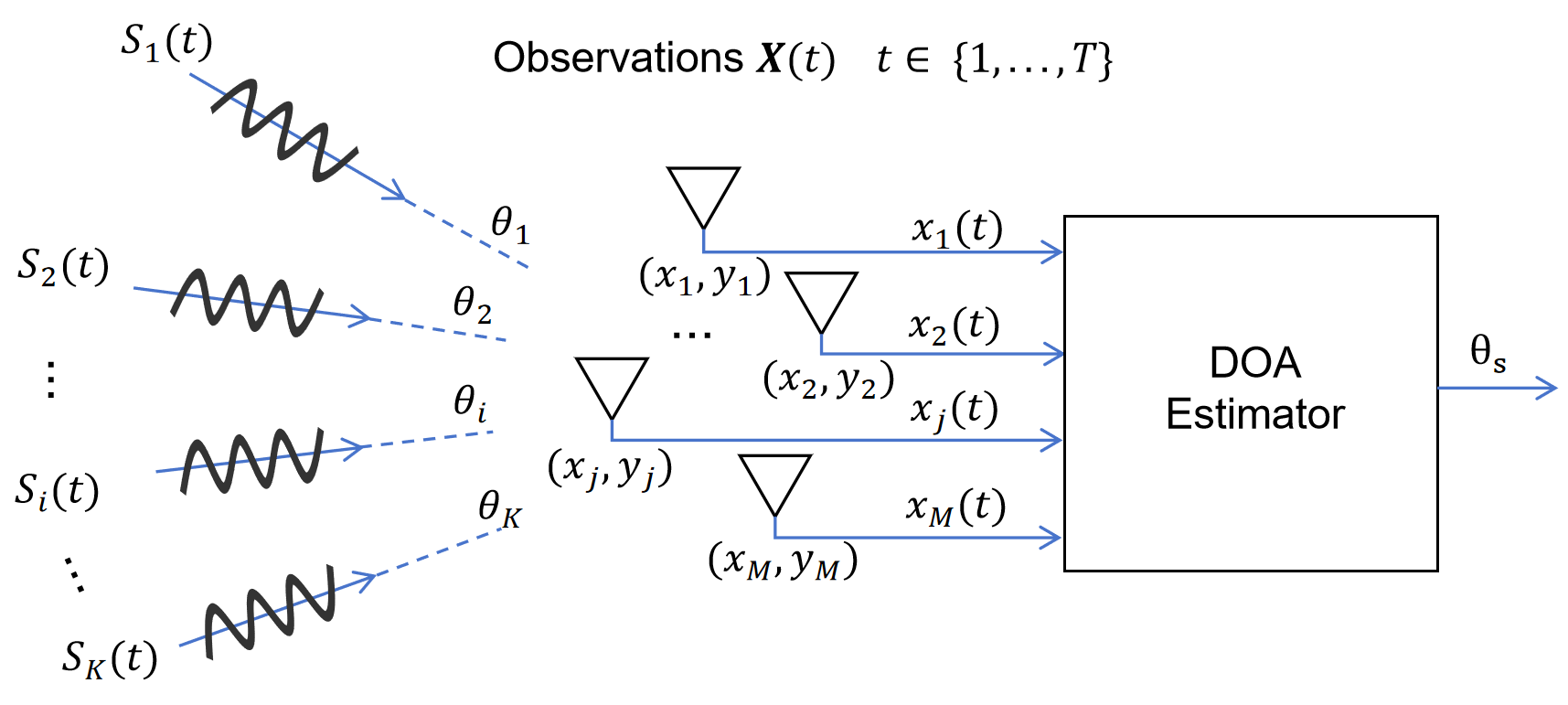}
	\caption{The process of direction-of-arrival estimation.}
	\label{fig:DOA_estiamtion_illustration}
\end{figure}

This DOA estimation problem inherently includes estimating the number of signal sources $K$. Furthermore, we specifically consider the extreme case where $T=1$, meaning the array snapshot data vector $\bm{X}(t)$ contains only one snapshot's worth of data. Figure \ref{fig:DOA_estiamtion_illustration} illustrates the process from array signal reception to direction-of-arrival estimation.

\subsection{Foundations of proposed algorithm}
\label{sec:foundations_of_proposed_algorithm}
Since the number of signals is typically limited in array processing, literature ~\cite{malioutov2005sparse} leverages the inherent sparsity of sources in the spatial domain to construct a sparse signal representation. Following the sparse characteristics, we utilize the spatial sparsity of the signal sources to represent the signal. Specifically, spatial directions containing no source signal are set to zero, while directions containing a source retain the corresponding source signal. Consequently, the signal reception model described in Formula~\eqref{eq:x=as+n} can be reformulated as:
\begin{equation}
\label{eq:x=as+n_complete}
\begin{bmatrix}
x_1(t) \\
x_2(t) \\
\vdots \\
x_M(t)
\end{bmatrix}=
\begin{bmatrix}
\bm{a}_{1}(\omega_0)...\bm{a}_{R}(\omega_0)
\end{bmatrix}
\begin{bmatrix}
s_1(t) \\
s_2(t) \\
\vdots \\
s_R(t)
\end{bmatrix}+
\begin{bmatrix}
n_1(t) \\
 \\
n_2(t) \\
\vdots \\
n_M(t)
\end{bmatrix}
\end{equation}
of which
\begin{equation}
    \label{eq:s_i(t)=0_s_i(t)=s_i(t)}
    \begin{cases}s_i(t)=0, & i \notin \widetilde{\theta}_s \cap \theta_s \\s_i(t)=s_i(t), & i \in \widetilde{\theta}_s \cap \theta_s
    \end{cases},
\end{equation}
where $\widetilde{\theta}_s$ is the set of all angular directions in space under precision $\delta$, $\theta_s$ is the set of all signal source directions in precision $\delta$, $R$ represents the length of set $\widetilde{\theta}_s$ ($R = |\widetilde{\theta_s}|=\frac{360}{\delta}$), $s_i(t)$ represents the signal function of time $t$ from the source in the $i$-th direction. At the moment of $t$, the spatial signal vector $\bm{S}(t)$ is sparse. Considering the discrete form, the signal reception model, i.e. Formula~\eqref{eq:x=as+n_complete}, can be represented by a matrix as:
\begin{equation}
\label{eq:X=A_complete S+N}
\bm{X}=\widetilde{\bm{A}}\bm{S}+\bm{N},
\end{equation}
here, we use $\widetilde{\bm{A}}$ to represent $\begin{bmatrix}
\bm{a}_{1}(\omega_0)...\bm{a}_{R}(\omega_0)
\end{bmatrix}$. Since the positional coordinates of the $M$ array elements, the propagation speed $c$, and the center frequency $\omega_0$ are known, $\widetilde{\bm{A}}$ is known. And $\bm{X}$ is the array received signal matrix of dimension $M \times T$, $\bm{S}$ is the signal matrix of dimension $R \times T$, $\bm{N}$ is the noise matrix of dimension $M \times T$, and $\widetilde{\bm{A}}$ is the array manifold matrix of dimension $M \times R$, which encompasses all spatial directions under the accuracy $\delta$. In this paper, $\widetilde{\bm{A}}$ is referred to as the complete array manifold matrix.

We consider the extreme case with $T=1$, where the received signal matrix $\bm{X}$ contains only data from a single snapshot. Under these conditions of utilizing spatial signal sparsity, discrete representation, and with $T=1$, the DOA estimation problem can be reformulated as:

\textit{Under the assumptions of a far-field, narrowband signal, ideal array conditions, and $\varphi = 0$ with $\theta_s^{'} = \theta_s$, in Formula~\eqref{eq:x=as+n_complete}, given the known complete array manifold matrix $\widetilde{\bm{A}}$ and the array received signal matrix $\bm{X}$ comprising only one snapshot, how to estimate the signal matrix $\bm{S}$, which according to Formula~\eqref{eq:s_i(t)=0_s_i(t)=s_i(t)}, the non-zero entries implicitly define the set $\theta_s$?} 

In the reformulated DOA estimation problem, the interpretations of the rows and columns for matrices $\bm{X}$, $\bm{\widetilde{A}}$, $\bm{S}$, and $\bm{N}$ are provided in Table \ref{tabel:interpretations_of_matrices_X_A_S_N}.

\begin{table*}
\caption{The interpretations of the rows and columns for matrices $\bm{X}$, $\bm{A}$, $\bm{S}$, and $\bm{N}$.}
\centering
\label{tabel:interpretations_of_matrices_X_A_S_N}
\begin{tabular}
{|l|l|l|l|l|}
\hline
Symbol & Explaination           & Shape & The interpretation of rows            & The interpretation of colums      \\ \hline
$\bm{X}$      & Array received signal matrix & $M \times 1$ & $M$ represents the array element domain & $1$ represents the current snapshot \\ \hline
$\bm{\widetilde{A}}$      & Complete array manifold matrix   & $M \times R$  & $M$ represents the array element domain & $R$ represents the spatial domain   \\ \hline
$\bm{S}$      &  Signal matrix & $R \times 1$  & $R$ represents the spatial domain & $1$ represents the current snapshot \\ \hline
$\bm{N}$     & Noise matrix  & $M \times 1$ & $M$ represents the array element domain & $1$ represents the current snapshot                             \\ \hline
\end{tabular}
\end{table*}

\subsection{Optimal Spatial Filter}
\label{subsec:Optimal Spatial Filter}
In section \ref{sec:foundations_of_proposed_algorithm}, we introduced the array signal model for sparse sources in space. As presented in Table \ref{tabel:interpretations_of_matrices_X_A_S_N}, the matrices $\bm{X}$, $\bm{\widetilde{A}}$, $\bm{S}$ and $\bm{N}$ reside in different domains: the spatial domain and the array element domain. To facilitate solving for the direction $\theta_s$ in the sparse matrix $\bm{S}$, we introduce the domain transformation matrix $\bm{B}$. Its purpose is to convert $\bm{X}$, $\bm{\widetilde{A}}$, and $\bm{N}$ from the array element domain to the spatial domain, thereby aligning them within the same spatial domain. This unification allows the DOA to be estimated within the spatial domain. The interpretation of the domain transformation matrix $\bm{B}$ is detailed in Table \ref{tabel:interpretations_of_matrices_B_BX_BA_BN}.

\begin{table*}
\caption{The interpretations of the rows and columns for matrices $\bm{B}$, $\bm{BX}$, $\bm{B\widetilde{A}}$, and $\bm{BN}$.}
\centering
\label{tabel:interpretations_of_matrices_B_BX_BA_BN}
\begin{tabular}
{|l|l|l|l|l|}
\hline
Symbol & Explaination           & Shape & The interpretation of rows            & The interpretation of colums      \\ \hline
$\bm{B}$      & Domain transformation matrix & $R \times M$ & $R$ represents the spatial domain & $M$ represents the array element domain \\ \hline
$\bm{BX}$      & Array received signal matrix in the spatial domain   & $R \times 1$  & $R$ represents the spatial domain & $1$ represents the current snapshot   \\ \hline
$\bm{B\widetilde{A}}$      &  Complete array manifold matrix in the spatial domain & $R \times R$  & $R$ represents the spatial domain & $R$ represents the spatial domain \\ \hline
$\bm{BN}$     & Noise matrix in the spatial domain  & $R \times 1$ & $R$ represents the spatial domain & $1$ represents the current snapshot                             \\ \hline
\end{tabular}
\end{table*}

\begin{table*}
\caption{The interpretations of the matrices $\bm{c_i}$ and $\bm{C}$.} 
\centering
\label{tabel:interpretations_of_matrices_C_ci}
\begin{tabular}
{|l|l|l|l|l|}
\hline
Symbol & Explaination           & Shape & The interpretation of rows            & The interpretation of colums      \\ \hline
$\bm{c_i}$      & Weight vectors for R directions in space   & $1 \times R$  & The $i$-th weight vector & $R$ spatial directions  \\ \hline
$\bm{C}$      & Weight matrix weighted in different directions of space & $R \times R$ & $R$ represents the spatial domain & $R$ represents the spatial domain \\ \hline
\end{tabular}
\end{table*}

Multiplying the domain transformation matrix $\bm{B}$ on both sides of Formula~\eqref{eq:x=as+n_complete} yields:
\begin{equation}
    \label{eq:BX=BAS+BN}
    \bm{BX} = \bm{B\widetilde{A}S} + \bm{BN},
\end{equation}
where the matrix $\bm{BX}$ represents the transformation of the array received signal matrix $\bm{X}$ from the array element domain to the spatial domain, $\bm{B\widetilde{A}}$ represents the transformation of the complete array manifold matrix $\bm{\widetilde{A}}$ from the array element domain to the spatial domain, and $\bm{BN}$ represents the transformation of the noise matrix $\bm{N}$ from the array element domain to the spatial domain. The interpretation of the matrix $\bm{BX}$, $\bm{B\widetilde{A}}$ and $\bm{BN}$ is detailed in Table \ref{tabel:interpretations_of_matrices_B_BX_BA_BN}.

Through the domain transformation matrix $\bm{B}$, we transformed the entire mathematical model corresponding to~\eqref{eq:X=A_complete S+N} into the spatial domain (Formula~\eqref{eq:BX=BAS+BN}). The advantage of this transformation is that it unifies the domains of variables, facilitating the solution for the spatial signal direction $\theta_s$. Domain transformation matrix $\bm{B}$ is also known as a spatial filter.

In Formula~\eqref{eq:BX=BAS+BN}, we pay particular attention to the matrix $\bm{B\widetilde{A}}$, where we define:
\begin{equation}
    \label{eq:BA=C}
    \bm{B\widetilde{A}}=\bm{C}
\end{equation}
of which
\begin{equation}
\bm{C}=
\begin{bmatrix}
\bm{c_1},\bm{c_2},\dots,\bm{c_R}
\end{bmatrix}^T,
\end{equation}
with
\begin{equation}
    \bm{c_i}=
\begin{bmatrix}
c_{i1},c_{i2},\dots,c_{iR}
\end{bmatrix}, i=1...R .
\end{equation}

Then Formula~\eqref{eq:BX=BAS+BN} becomes:
\begin{equation}
    \label{eq:BX=CS+BN}
    \bm{BX} = \bm{CS} + \bm{BN}    
\end{equation}

Regarding the values in each row $\bm{c_i}$ of matrix $\bm{C}$ as the weights corresponding to $R$ spatial directions in space, each of these $R$ spatial directions will then correspond to $R$ weights. The meanings of $\bm{c_i}$ and $\bm{C}$ are as shown in Table~\ref{tabel:interpretations_of_matrices_C_ci}.

If $\bm{C}$ is the identity matrix $\bm{E}$, then the meaning of $\bm{C}$ is that the weight vector $\bm{c_i}$ for the $i$-th direction is 1, while the weights for other directions are 0 (if the weights for other directions were non-zero, interference from those directions would be introduced). This way, $\bm{c_i}$ focuses on the current $i$-th space. With $R$ distinct weights vector $\bm{c_i}$, the role of $\bm{C}$ is to focus on $R$ different directions respectively. The advantage of this is that the original signal of the current spatial direction is obtained by focusing on a single spatial direction. Then, the Formula~\eqref{eq:BX=CS+BN} becomes:
\begin{equation}
    \label{eq:BX=S+BN}
    \bm{BX} = \bm{S}+\bm{BN}.
\end{equation}
If matrix $\bm{B}$ also satisfies $\bm{BN}=0$, then after transforming the noise $\bm{N}$ from the array element domain to the spatial domain using $\bm{B}$, the noise will be dissipated. Eventually, formula~\eqref{eq:BX=S+BN} becomes $\bm{BX=S}$, where $\bm{C=E}$ and $\bm{BN=0}$. If we can obtain such a matrix $\bm{B}$, then $\bm{S}$ can be computed by $\bm{BX}$ ($\bm{X}$ is known). According to Formula~\eqref{eq:s_i(t)=0_s_i(t)=s_i(t)} and the sparsity of $\bm{S}$, $\theta_s$ can then be determined (corresponding to the non-zero directions in $\bm{S}$). This spatial filter $\bm{B}$ is what we want; it satisfies the conditions of the optimal spatial filter that will be defined next:
\begin{definition}[Optimal spatial filter]
\label{def:optimal spatial filter}
If the spatial filter matrix $\bm{B}$ satisfies Formula~\eqref{eq:BA=C} and~\eqref{eq:BX=CS+BN} and also satisfies the following formula:
\begin{equation}
    \label{eq:BX=S}
    \bm{BX = S},
\end{equation}
then we refer to the spatial filter matrix $\bm{B}$ as the optimal spatial filter, denoted by $\bm{B^*}$.
\end{definition}

One sufficient condition for spatial filter $\bm{B}$ to satisfy Definition~\ref{def:optimal spatial filter} is:
\begin{numcases}{}
\bm{B\widetilde{A}S} = \bm{S} \label{eq:BAS=S} \\
\bm{BN} = \bm{0} \label{eq:BN=0}
\end{numcases}
The process described by Formula~\eqref{eq:BAS=S} is referred to as spatial signal focusing, and the process described by Formula~\eqref{eq:BN=0} is referred to as noise dispersion.

When we find a spatial filter $\bm{B}$ that satisfies both Formula~\eqref{eq:BAS=S} and Formula~\eqref{eq:BN=0}, then according to sufficient conditions, $\bm{B}$ must satisfy Formula~\eqref{eq:BX=S} to become the optimal spatial filter. Once we find the optimal spatial filter, we can calculate $\bm{S}$ according to Formula~\eqref{eq:BX=S}, and then obtain $\theta_s$ based on Formula~\eqref{eq:s_i(t)=0_s_i(t)=s_i(t)} and the sparsity of $\bm{S}$.

$\bm{B^*X}$ contains the directions of the sparse signal matrix $\bm{S}$ in space and the source signals $\bm{S_i}$ in each direction. Therefore, solving the direction of arrival of signals in space can be transformed into solving the optimal spatial filter $\bm{B^*}$, satisfying Formula~\eqref{eq:BA=C},~\eqref{eq:BX=CS+BN}, and~\eqref{eq:BX=S}.

The spatial signal focusing and noise suppression algorithm proposed in this paper centers on the sufficient conditions: Formula~\eqref{eq:BAS=S} and Formula~\eqref{eq:BN=0} to seek the optimal filter $\bm{B^*}$ so as to solve $\theta_s$.

\section{Spatial Signal Focusing and Noise Suppression Algorithm}
\label{sec:Spatial Signal Focusing and Noise Suppression Algorithm}

In this section, we will introduce our proposed algorithm. We consider the general case that the number of array elements $M$ is less than the number of spatial directions $R$.

\subsection{The existence of the optimal spatial filter}

In this section, we will discuss the existence of the optimal filter $\bm{B^*}$. We first assume the complete array manifold matrix $\bm{\widetilde{A}}$ and the noise $\bm{N}$ are linearly independent; otherwise, Formulas~\eqref{eq:BAS=S} and~\eqref{eq:BN=0} may not hold simultaneously.

Considering Formulas~\eqref{eq:BAS=S} and~\eqref{eq:BN=0}, the degree of freedom of each row $\bm{b_i}$ of the spatial filter $\bm{B}$ is $M-K-1$, which is limited by the noise matrix $\bm{N}$ and the $K$ signal sources in the signal matrix $\bm{S}$. In order to satisfy Formulas~\eqref{eq:BAS=S} and~\eqref{eq:BN=0}, the optimal spatial filter $\bm{B^*}$ exists when $M-K-1 \geq 0$, and does not exist when $M-K-1 < 0$. It is worth noting that when $M-K-1 > 0$, the optimal spatial filter $\bm{B^*}$ is not unique because the degrees of freedom of $\bm{B}$ are not fully restricted; When $M-K-1 = 0$, the optimal spatial filter $\bm{B^*}$ is unique. We consider the case where the optimal spatial filter $\bm{B^*}$ exists, assuming that the number of signal sources $K$ in space is less than the number of array elements $M$ to satisfy $M-K-1 \geq 0$ . In this case, the problem of solving DOA can be reorganized as:

\textit{Under the assumptions of a far-field, narrowband signal, ideal array conditions, $\varphi = 0$ with $\theta_s^{'} = \theta_s$, $K < M$ and $M < R = \frac{360}{\delta}$, in Formula~\eqref{eq:x=as+n_complete}, given the known complete array manifold matrix $\widetilde{\bm{A}}$ and the array received signal matrix $\bm{X}$ comprising only one snapshot, how to estimate the optimal spatial filter $\bm{B^*}$ to satisfy both Formula~\eqref{eq:BAS=S} and Formula~\eqref{eq:BN=0} simultaneously, which can obtain the signal matrix $\bm{S}$ by using Formula~\eqref{eq:BX=S} and thus $\theta_s$ can be obtained ?}

Because the noise matrix $\bm{N}$ does not have sparsity and has randomness and unknowability, it is difficult to solve for $\bm{B^*}$ that satisfies Formula~\eqref{eq:BN=0}. Because our goal is to solve $\theta_s$ based on the sparsity of $\bm{S}$, we focus more on Formula~\eqref{eq:BAS=S}, so we can relax the conditions of Formula~\eqref{eq:BN=0} to approximate it, that is, we solve for the approximate $\bm{B^{'}}$ of the optimal spatial filter $\bm{B^*}$, which satisfies the following formula:
\begin{equation}
    \label{eq:BN approx 0}
    \begin{cases}
    \bm{B^{'}\widetilde{A}S} = \bm{S} \\
\bm{B^{'}N} \approx \bm{0}
    \end{cases},
\end{equation}
at this point, $\bm{B^{'}X}\approx \bm{S}$. We refer to the process of $\bm{B^{'}N} \approx 0$ as noise suppression. So the goal of the \textit{Spatial Signal Focusing and Noise Suppression} algorithm is to solve for the spatial filter $\bm{B^{'}}$ that satisfies Formula~\eqref{eq:BN approx 0}, in order to approximate the optimal spatial filter $\bm{B^*}$ as much as possible.

For Formula~\eqref{eq:BN approx 0}, we perform an approximate estimation of the optimal spatial filter $\bm{B^*}$ in two steps. Firstly, we solve the spatial filter $\bm{B}$ that satisfies the following formula to obtain a preliminary estimation of $\bm{B^{'}}$:
\begin{equation}
    \label{eq:BAS&BN approx 0}
    \begin{cases}
    \bm{B\widetilde{A}S} \approx \bm{S} \\
\bm{BN} \approx \bm{0}
    \end{cases}.
\end{equation}

Then, we eliminate the mutual interference of spatial signals to obtain Formula~\eqref{eq:BN approx 0}. In the next two sections, we will provide a detailed introduction to these two steps.

\subsection{Preliminary estimation}
If $\bm{BA}$ can be made equal to the identity matrix $\bm{E}$, then Formula~\eqref{eq:BAS=S} naturally satisfies. However, since the number of array elements $M$ is less than the number of spatial directions $R$, the mathematically complete array manifold matrix $\bm{\widetilde{A}}$ does not have a left inverse $\bm{B}$ such that $\bm{BA=E}$. The advantage of the identity matrix $\bm{E}$ is that it can focus on signals in the current spatial direction without introducing signals from other directions. In order to enable spatial filter $\bm{B}$ to focus on signals in the current spatial direction and introduce as few signals from other directions as possible, we expect the $i$-th spatial weight vector to have a weight $c_{ii}$ equal to 1 for the i-th direction, while the weight $c_{ij}~(i \neq j)$ for other directions approaches 0. That is to say:

\begin{equation}
    \label{eq:c_ii=biai}
    \begin{cases}
    c_{ii} = \bm{b_i}\bm{a_i} = 1,~i=1 \dots R \\
    c_{ij} = \bm{b_i}\bm{a_j}\approx0,~j=1 \dots R,~j \neq i,
    \end{cases}
\end{equation}
where $\bm{b_i}$ represents the $i$-th row vector in the spatial filter $\bm{B}$, and $\bm{a_j}$ represents the $j$-th column vector in the complete array manifold matrix $\bm{\widetilde{A}}$. In addition, we also expect that the spatial filter $\bm{B}$ can suppress noise, so for $\bm{b_i}$:
\begin{equation}
    \label{eq:biN}
    \bm{b_i}\bm{N}\approx0,~i=1 \dots R.
\end{equation}

For Formula~\eqref{eq:c_ii=biai} and~\eqref{eq:biN}, we take their energy, which is:
\begin{equation}
    \label{eq:|biai|^2=1}
    \begin{cases}
    |\bm{b_i}\bm{a_i}|^2  =1,~i=1 \dots R \\
    |\bm{b_i}\bm{a_j}|^2\approx0,~j=1 \dots R, j \neq i\\
    |\bm{b_i}\bm{N}|^2\approx0,~i=1 \dots R \\
    \end{cases}
\end{equation}

The $||$ symbol represents taking the modulus of each element in matrices or vectors. $|\bm{BX}|^2$ represents the spatial energy spectrum. According to Formula~\eqref{eq:BX=BAS+BN}, we have:
\begin{equation}
    |\bm{BX}|^2 = |\bm{B\widetilde{A}S} + \bm{BN}|^2 \leq |\bm{B\widetilde{A}S}|^2 + |\bm{BN}|^2
\end{equation}

For each row $\bm{b_i}$ in spatial filter $\bm{B}$, there are:
\begin{equation}
\begin{aligned}
    |\bm{b_iX}|^2 = &|\bm{b_i\widetilde{A}S} + \bm{b_iN}|^2 \\
    \leq &|\bm{b_i\widetilde{A}S}|^2 + |\bm{b_iN}|^2\\
    = &|\bm{b_ia_i}s_i+\sum_{i \neq j}\bm{b_ia_j}s_j|^2 + |\bm{b_iN}|^2\\
    \leq &|\bm{b_ia_i}|^2|s_i|^2+\sum_{i \neq j}(|\bm{b_ia_j}|^2|s_j|^2) + |\bm{b_iN}|^2
\end{aligned}
\end{equation}

If the Formulas~\eqref{eq:|biai|^2=1} are satisfied, then:
\begin{equation}
\label{eq:s_i^2<=b_iX<=}
|s_i|^2 \leq |\bm{b_iX}|^2 \leq |s_i|^2 + \epsilon,
\end{equation}
where $\epsilon \approx 0$. This indicates that each row $\bm{b_i}$ in the spatial filter $\bm{B}$ filters the signal for the $i$-th spatial direction and reflects its energy magnitude. Consequently, $|\bm{BX}|^2 \approx|\bm{S}|^2$, revealing the energy spectrum of the spatial source signals.

For $|\bm{b_ia_j}|^2$ and $|\bm{b_iN}|^2$, the following transformation has been applied:
\begin{equation}
\begin{aligned}
    |\bm{b_ia_j}|^2 = &|b_{i1}a_{j1}+\dots+b_{iM}a_{jM}|^2\\
    \leq &|b_{i1}a_{j1}|^2+\dots+|b_{iM}a_{jM}|^2\\
    =&|b_{i1}|^2|a_{j1}|^2+\dots+|b_{iM}|^2|a_{jM}|^2
\end{aligned}
\end{equation}
and
\begin{equation}
\begin{aligned}
    |\bm{b_iN}|^2 = &|b_{i1}n_1+\dots+b_{iM}n_M|^2\\
    \leq &|b_{i1}n_1|^2+\dots+|b_{iM}n_M|^2\\
    =&|b_{i1}|^2|n_1|^2+\dots+|b_{iM}|^2|n_M|^2
\end{aligned}
\end{equation}
where $a_{ij}$ represents the $j$-th element in $\bm{a_i}$, $b_{ij}$ represents the $j$-th element in $\bm{b_i}$, $n_i$ represents the $i$-th element in $\bm{N}$.

If $|b_{i1}|^2=...=|b_{iM}|^2$, then:
\begin{equation}
    |\bm{b_ia_j}|^2 \leq |b_{i1}|^2(|a_{j1}|^2+\dots+|a_{jM}|^2)
\end{equation}
\begin{equation}
    |\bm{b_iN}|^2 \leq |b_{i1}|^2(|n_{1}|^2+\dots+|n_M|^2)
\end{equation}

At this point, the closer $|b_{i1}|^2$ approaches 0, the closer $|\bm{b_i}N|^2$ and $|\bm{b_ia_j}|^2 $ approach 0. If it also satisfies $|\bm{b_ia_i}|^2 = 1$, according to Formula~\eqref{eq:s_i^2<=b_iX<=}, then $|\bm{b_i\widetilde{A}S}|^2$ becomes closer to $|s_i|^2$, and $|\bm{BX}|^2$ becomes closer to $|\bm{S}|^2$.

In summary, in order to simultaneously meet the requirements of spatial signal focusing and noise suppression, we have the following optimization formula:
\begin{equation}
\label{optim:min |b_i1|}
\begin{aligned}
 \mathrm{min}~&|b_{i1}|^{2} \\
 \mathrm{s.t.}~&|b_{i1}|^2=\dots=|b_{iM}|^2\\
 &|\bm{b_ia_i}|^2=1
\end{aligned}\end{equation}
The problem above is a quadratic programming problem with equality constraints, and its optimal solution is:
\begin{equation}
    \label{eq:b_ij=a_ij/sum}
     b_{ij} = \frac{a_{ij}.conj()}{|a_{ij}|\sum_{j=1}^M |a_{ij}|},i=1\dots R, j=1\dots M.
\end{equation}

so, we can obtain $\bm{b_i}$ by:
\begin{equation}
     \bm{b_i} = [b_{i1},\dots,b_{ij},\dots,b_{iM}]
\end{equation}

We combine the obtained $\bm{b_i}$ in row order to form a spatial filter $\bm{B}$ and derive the preliminarily estimated spatial filter $\bm{B^{'}}$ that satisfies Formula~\eqref{eq:BAS&BN approx 0}, thereby completing the first step of the spatial signal focusing and noise suppression algorithm.

The approximation of the optimal spatial filter obtained above cannot guarantee that the weight $c_{ij}$ in other spatial directions where signals exist is zero, resulting in mutual interference between signals in space. So, in the next section, we will introduce that based on the approximation of the optimal spatial filter obtained as the initial value which satisfies Formula~\eqref{eq:BAS&BN approx 0}, we iteratively optimize $\bm{B}$ by seeking the maximum value under certain conditions to obtain $\bm{B^{'}}$ that satisfies $\bm{B^{'}\widetilde{A}S=S}$ to eliminate the mutual interference of spatial signals, that is, to satisfy Formula~\eqref{eq:BN approx 0}.

\subsection{Eliminate mutual interference of spatial signals}
\label{subsection:Eliminate mutual interference of spatial signals}
Although the weight matrix $\bm{C=B\widetilde{A}}$ obtained from the spatial filter $\bm{B}$ through preliminary estimation in the previous section did not completely eliminate the interference from other directions where signals exist in space, this weight matrix $\bm{C}$ effectively enhances the signal from the current spatial direction. Therefore, we consider that the direction corresponding to the maximum value of the spatial spectrum $\bm{|BX|=|B\widetilde{A}S+BN|}$ holds certain significance.

In order to separate signals in space, we expect
\begin{equation}
    \frac{|c_{ii}|}{Max_{j \neq i,j=1\dots R}|c_{ij}|}
\end{equation}
to be as large as possible, so as to focus on signals in the current spatial direction and minimize interference from other signal directions. We denote
\begin{equation}
    \frac{1}{q_i} = \frac{|c_{ii}|}{Max_{j \neq i,j=1\dots R}|c_{ij}|}.
\end{equation}
Because we set $|c_{ii}| = |\bm{b_ia_i}| = 1$, therefore
\begin{equation}
    q_i = Max_{j \neq i,j=1\dots R}|c_{ij}|.
\end{equation}

Let
\begin{equation}
    q=Max~q_i.
\end{equation}

The component in the $i$-th direction of the spatial spectrum $|\bm{BX}|$ can be represented as
\begin{equation}
    \label{eq:|b_iX|=sumc_ijs_j}
    |\bm{b_iX}|=|\sum_{j=1}^Rc_{ij}s_j+\bm{b_iN}|.
\end{equation}

According to Formula~\eqref{eq:s_i(t)=0_s_i(t)=s_i(t)}, $\bm{S}$ has sparsity, Formula~\eqref{eq:|b_iX|=sumc_ijs_j} can be written as
\begin{equation}
    |\bm{b_iX}|=|\sum_{j\times\delta \in \theta_s}c_{ij}s_j+\bm{b_iN}|.
\end{equation}

We will discuss two situations where $i\times\delta \notin \theta_s$ and $i\times\delta \in \theta_s$. When the domain of $i$ satisfies $i\times\delta \notin \theta_s$, we have:
\begin{equation}
\begin{aligned}
    |\bm{b_iX}| = &|\sum_{j\times\delta \in \theta_s}c_{ij}s_j+\bm{b_iN}| \\
    \leq&|\sum_{j\times\delta \in \theta_s}qs_j+\bm{b_iN}|\\
     \leq&|qKMax_{j\times\delta \in \theta_s}s_j+\bm{b_iN}|,i\times\delta \notin \theta_s
\end{aligned}
\end{equation}

When the domain of $i$ satisfies $i\times\delta \in \theta_s$ and $i=argmax_i |\bm{b_iX}|$, we have:
\begin{equation}
    \begin{aligned}
    |\bm{b_iX}| = &|\sum_{j\times\delta \in \theta_s}c_{ij}s_j+\bm{b_iN}| \\
    &\geq|Max_{j\times\delta \in \theta_s}s_j+\bm{b_iN}|,\\
    &i\times\delta \in \theta_s, i=argmax_i |\bm{b_iX}|
    \end{aligned}
\end{equation}

If the following inequality holds true:
{\small
\begin{equation}
\label{eq:|qKMax+bN|<=|Max+bN|}
\begin{aligned}
    &|qKMax_{j\times\delta \in \theta_s}s_j+\bm{b_iN}|,i\times\delta \notin \theta_s\\
    &\leq |Max_{j\times\delta \in \theta_s}s_j+\bm{b_iN}|,i\times\delta \in \theta_s,i=argmax_i |\bm{b_iX}|,
\end{aligned}
\end{equation}
}there must be a signal in the spatial direction corresponding to the maximum value of the spatial spectrum $|\bm{BX}|$. The advantage of this is that regardless of the shape of the spatial spectrum $|\bm{BX}|$, its maximum value corresponds to the direction of the signal. Subsequently, the spatial filter $\bm{B}$ can be iterated based on the current obtained signal direction, so that after each iteration, the maximum value in the spatial spectrum $|\bm{BX}|$ can be used to determine the next signal direction.

Because we have $\bm{BN}\approx\bm{0}$, we assume that each component in $\bm{BN}$ is approximately equal; then Formula~\eqref{eq:|qKMax+bN|<=|Max+bN|} can be written as:
\begin{equation}
\label{eq:|qKMax|<=|Max|}
    |qKMax_{j\times\delta \in \theta_s}s_j|\leq |Max_{j\times\delta \in \theta_s}s_j|.
\end{equation}

When $q<\frac{1}{K}$, Formula~\eqref{eq:|qKMax|<=|Max|} holds, and the spatial direction corresponding to the maximum value in the spatial spectrum $|\bm{BX}|$ represents the signal direction.

It should be noted that the value of $q$ depends on the weight matrix $\bm{C}$, while the weight matrix $\bm{C}$ is determined by the spatial filter $\bm{B}$ and the complete array manifold matrix $\bm{\widetilde{A}}$. According to Formula~\eqref{eq:b_ij=a_ij/sum}, the spatial filter $\bm{B}$ is itself determined by the complete array manifold matrix $\bm{\widetilde{A}}$. The complete array manifold matrix $\bm{\widetilde{A}}$, in turn, is determined by the array arrangement and the number of array elements. Therefore, the value of $q$ is ultimately determined by the array arrangement and the number of array elements. In subsequent subsections, we explore the influence of different array arrangements and numbers of array elements on $q$. In order to separate the signal, in the subsequent, we assume that our array satisfies the condition $q<\frac{1}{K}$, which means that the direction of the maximum value in the spatial spectrum corresponds to the direction of the signal.

We denote the spatial direction corresponding to the maximum value in the current spatial spectrum $|\bm{BX}|$ as:
\begin{equation}
    i^*=argmax~\bm{|b_iX|}.
\end{equation}
According to the assumption of $q< \frac{1}{K}$, there is a signal in the direction $i^*$ corresponding to the maximum value in the spatial spectrum. When we know that there is a signal in a certain direction, we can eliminate the influence of the signal on other spatial directions, thereby reducing the mutual interference between signals. That is, we can make the weight $|c_{i^*i^*}| = 0$ in the $i^* $ spatial direction to eliminate the signal in that spatial direction.

For optimization Formula~\eqref{optim:min |b_i1|}, we can use $|c_{i^*i^*}| = 0$ as a constraint to optimize the solution of spatial filter B and eliminate the interference of known signals. We have the following optimization formula:
\begin{equation}
\label{optim:min |b_i1| with c_i^*i^*=0}
\begin{aligned}
 \mathrm{min}~&|b_{i1}|^{2} \\
 \mathrm{s.t.}~&|b_{i1}|^2=\dots=|b_{iM}|^2\\
 &|\bm{b_ia_i}|^2=1,~i\neq i^*\\
 &|\bm{b_{i^*}a_{i^*}}|^2=0
\end{aligned}\end{equation}

It should be noted that the constraints in optimization Formula~\eqref{optim:min |b_i1| with c_i^*i^*=0} cannot be mutually satisfied, mainly because the constraint condition $|b_{i1}|^2=\dots=|b_{iM}|^2$ is quite strict. In order to relax $|b_{i1}|^2=\dots=|b_{iM}|^2$ condition, while satisfying $|b_{i1}|^2,\dots,|b_{iM}|^2$ can approximate the condition $|b_{i1}|^2=\dots=|b_{iM}|^2$ and $min~|b_{i1}|^2$, we hope that $|b_{i1}|^2,\dots,|b_{iM}|^2$ has a small variance and mean, i.e. $min~(|b_{i1}|^2+\dots+|b_{im}|^2)$. So the optimization formula for updating spatial filter $\bm{B}$ by eliminating known signal interference is:
\begin{equation}
\begin{aligned}
 \mathrm{min}~&|b_{i1}|^2+\dots+|b_{im}|^2 \\
 \mathrm{s.t.}~&|\bm{b_ia_i}|^2=1,~i\neq i^*\\
 &|\bm{b_{i^*}a_{i^*}}|^2=0
\end{aligned}\end{equation}

At this point, the signal in the $i^*$ spatial direction is eliminated on the basis of the original spatial filter $\bm{B}$, and we obtain a new spatial filter $\bm{B}^{new}$. The new spatial filter $\bm{B}^{new}$ obtains a new spatial spectrum $|\bm{B}^{new}X|$. If the $q$ (also denoted $q^{new}$) calculated by $\bm{B}^{new}$ is less than $\frac{1}{K-1}$ and $k-1>0$ (the reason for $K-1$ is that one of the known signal directions is masked), then the maximum value of the newly obtained spatial spectrum $|\bm{B}^{new}X|$ also corresponds to a signal, and the new $i^*=argmax ~\bm{b_i}^{new}\bm{X}$. The above steps can iteratively add constraints based on the direction of the maximum value obtained in each iteration until the maximum value in the spatial spectrum $|\bm{BX}|$ is less than a certain threshold (indicating that the signal is completely eliminated and only noise remains) or reaches the maximum number of iterations $I$ ($I<M$, because $K<M$), then it is considered that all spatial directions where signals exist have been found.

The set of signal directions that need to be eliminated in the $t$-th iteration is denoted as $\theta_{t}$. $\theta_0$ represents an empty set. The optimization formula for spatial filter $\bm{B}$ in the $t$-th iteration to eliminate the signal direction set $\theta_{t}$ is:
\begin{equation}
\label{optim:min b_i1^2+...+b_im^2}
\begin{aligned}
 \mathrm{min}~&|b_{i1}|^2+\dots+|b_{im}|^2 \\
 \mathrm{s.t.}~&|\bm{b_ia_i}|^2=1,~i\notin \theta_{t}\\
 &|\bm{b_{i^*}a_{i^*}}|^2=0, i^* \in \theta_{t}
\end{aligned}\end{equation}
Its optimal solution is:
\begin{equation}
    \label{eq:b_i=a_iHQQH/|a_iHaQ|2}
    \bm{b_i} = \frac{\bm{a_i}^H\bm{QQ}^H}{|\bm{{a_i}}^H\bm{Q}|^2}
\end{equation}
with
\begin{equation}
    \bm{Q} = \mathrm{Orthonormal~Basis}(\mathcal{N}({A^{'}}^H)),
\end{equation}
where $A^{'}$ represents a matrix composed of $\bm{a_{i^*}}$ ($i^* \in \theta_{t}$) column vectors; $\mathcal{N}$ represents the null space of a matrix; $H$ represents conjugation.

After $I$ iterations, the final result is $\theta_I$ ($|\theta_I|<M$), which includes the spatial direction of the signal. Because $|\theta_I|$ may be greater than $K$, it may contain spatial directions where there is no signal present. We need to obtain $\theta_s$ from the set $\theta_I$.

We need to construct a spatial filter $\bm{B}$ based on the obtained $\theta_I$, so that the weight matrix $\bm{C}$ it obtains can highlight the signal in the current spatial direction and make the spatial weights of other possible signals zero, thereby eliminating interference in other signal spatial directions and minimizing $|b_{i1}|^2+\dots+|b_{im}|^2$ to suppress noise. The optimization formula for the spatial filter $\bm{B^{'}}$ that we ultimately obtain satisfies Formula~\eqref{eq:BN approx 0} is:
\begin{equation}
\label{optim:min b_i1^2+...+b_im^2_final}
\begin{aligned}
 \mathrm{min}~&|b_{i1}|^2+\dots+|b_{im}|^2 \\
 \mathrm{s.t.}~&|\bm{b_ia_i}|^2=1\\
 &|\bm{b_{i^*}a_{i^*}}|^2=0, i^* \in \theta_{I}-{\{i\}}
\end{aligned}\end{equation}
Its optimal solution form is the same as optimization Formula~\eqref{optim:min b_i1^2+...+b_im^2}.

When we obtain the spatial filter $\bm{B^{'}}$, we can obtain the spatial energy spectrum $|\bm{B^{'}X}|^2$ according to Formulas~\eqref{eq:BN approx 0} and~\eqref{eq:BX=BAS+BN}, and thus obtain $\theta_s$ based on the sparsity of the spatial signal. We provide the complete process of spatial signal focusing and noise suppression algorithms in Algorithm~\ref{alg:spatial signal focusing and noise suppression}, and provide examples in the Appendix for a single snapshot of 16 elements in a two-dimensional array.

All the discussions above are based on implementations under a single snapshot, without assuming signal non-coherence. Additionally, they do not require prior knowledge of the number of signals. The derived spatial filter $\bm{B}$ only depends on the complete array manifold matrix $\bm{\widetilde{A}}$; thus, the spatial filter $\bm{B}$ can be precomputed prior to signal processing.

\begin{algorithm}
\renewcommand{\algorithmicrequire}{\textbf{Input:}}
\renewcommand{\algorithmicensure}{\textbf{Output:}}
\caption{Proposed algorithm for DOA}\label{alg:spatial signal focusing and noise suppression}
\begin{algorithmic}[1]
\REQUIRE  $\bm{X}$ and $\bm{\widetilde{A}}$;
\ENSURE   $\theta_s$;
\STATE Initialize $\bm{B}$ through Formulas~\eqref{optim:min |b_i1|} and~\eqref{eq:b_ij=a_ij/sum};
\STATE Set threshold $P$ or maximum iteration count $I$;
\STATE Assuming $q<\frac{1}{K}$ holds true;
\STATE Initialize $t=1$;~$\theta_0=\emptyset$;~$i^*=argmax_i~|\bm{b_iX}|$;
\WHILE{$t\leq I$~or~$|\bm{b_{i^*}X}|<P$}
\STATE $\theta_t = \theta_{t-1} + \{i^*\}$
\STATE Update $\bm{B}$ through Formulas~\eqref{optim:min b_i1^2+...+b_im^2} and~\eqref{eq:b_i=a_iHQQH/|a_iHaQ|2};
\STATE $K=K-1$;~Assuming $q^{new}<\frac{1}{K}$ holds true;
\STATE $i^*=argmax_i~|\bm{b_iX}|$;
\STATE $t=t+1$
\ENDWHILE
\STATE Get $\theta_I~(\theta_I=\theta_{t-1})$;~Obtain $\bm{B^{'}}$ through Formula~\eqref{optim:min b_i1^2+...+b_im^2_final};
\STATE Calculate the spatial energy spectrum $|\bm{B^{'}X}|^2$;
\STATE Obtain $\theta_s$ from $|\bm{B^{'}X}|^2$ based on the sparsity of the signal.
\end{algorithmic}
\end{algorithm}

\subsection{Multiple snapshots}
Considering the case of multiple snapshot counts, the dimension of the array signal receiving matrix $\bm{X}$ is $M \times T$, where $T$ represents the snapshot count. We use $\bm{X}_i$ to represent the $i$-th snapshot in $\bm{X}$, so $\bm{X}_i$ has a dimension of $M \times 1$. In the case of multiple snapshots, we separately calculate the $|\bm{BX}_i | ^ 2$ spatial energy spectrum for each single snapshot, and then average the spatial energy spectrum for each snapshot to obtain the spatial energy spectrum for multiple snapshots, which is:
\begin{equation}
    \overline{|\bm{BX}|^2} = \frac{\sum_{i=1}^T|\bm{BX}_i|^2}{T}.
\end{equation}

\section{Effects of Array on Algorithm Performance}
\label{sec:Effects of Array on Algorithm Performance}
In the proposed algorithm, the solution of spatial filter $\bm{B}$ relies only on the complete array manifold matrix $\bm{\widetilde{A}}$ and $\bm{\widetilde{A}}$ is affected by the number of array elements, array arrangement, and array aperture. So in this section, we will explore the impact of these array factors on the performance of the algorithm. We assume a narrowband signal propagation speed $c=1500~m/s$ and a frequency $f$ of $100~Hz$. We scale the number of elements in the array from 8 to 256, and the aperture of the array from 0 to 10000 meters. The arrangement of the array is selected from two-dimensional uniform random distribution array, two-dimensional normal random distribution array, two-dimensional uniform circular array, two-dimensional uniform concentric circular array, two-dimensional spiral array, and one-dimensional uniform linear array. We recommend readers refer to the Appendix for detailed arrangement of these arrays.

\subsection{Effects of array on $q$}
\label{subsec:effects of array on q}
In the proposed algorithm, a core idea is that there is a signal in the direction corresponding to the maximum value of the spatial spectrum $|\bm{BX}|$. In section~\ref{subsection:Eliminate mutual interference of spatial signals}, we provide the condition that satisfies this idea: $q<\frac{1}{K}$. Because $q$ is determined by the array, we assume that the array satisfies this condition. However, it should be noted that condition $q<\frac{1}{K}$ is a sufficient condition for the presence of signals in the direction corresponding to the maximum value of the spatial spectrum $|\bm{BX}|$, rather than a necessary condition. That is, even if condition $q<\frac{1}{K}$ is not met, there may still be signals in the direction corresponding to the maximum value of the spatial spectrum $|\bm{BX}|$.

\begin{figure}[t] 
\centering 
\includegraphics[width=0.48\textwidth]{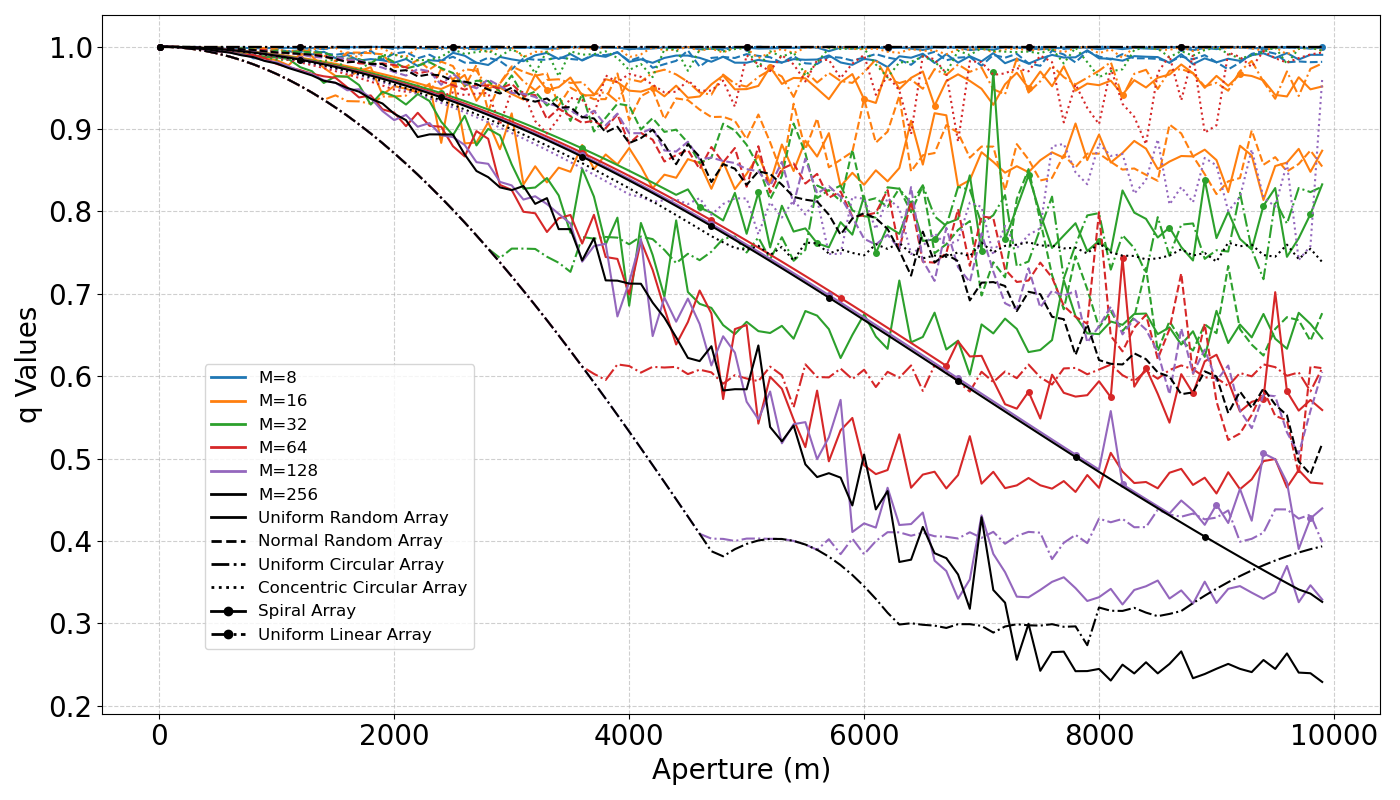} 
\caption{Effects of array on $q$.}  
\label{fig:effect_of_array_on_q} 
\end{figure}

Figure~\ref{fig:effect_of_array_on_q} shows the effect of the array on $q$ under different numbers of elements, array arrangements, and array apertures. From Figure~\ref{fig:effect_of_array_on_q}, we observe that $q<\frac{1}{K}$ is difficult to satisfy in a one-dimensional uniform linear array, which is not suitable for our proposed algorithm. In a two-dimensional array, we found that as the array aperture and number of elements increase, the value of $q$ becomes smaller, and the assumption that there is a signal in the direction corresponding to the maximum value of the spatial spectrum $|\bm{BX}|$ becomes easier to satisfy. Especially, the arrangement of two-dimensional uniform circular arrays and two-dimensional uniform random arrays can make $q$ have smaller values. According to the results in Figure~\ref{fig:effect_of_array_on_q}, in order to satisfy the core idea that there is a signal in the direction corresponding to the maximum value of the spatial spectrum $|\bm{BX}|$, we apply the proposed algorithm to a large aperture two-dimensional array.


\subsection{Effects of array on spatial signal focusing}
In our proposed algorithm, focusing on spatial signals is an important step, which is related to the weight matrix $\bm{C}$. The weight matrix $\bm{C}$ is related to the spatial filter $\bm{B}$ and the complete array manifold matrix $\bm{\widetilde{A}}$, and is calculated using formula~\eqref{eq:BA=C}. The solution of spatial filter $\bm{B}$ only relies on the complete array manifold matrix $\bm{\widetilde{A}}$, so the weight matrix $\bm{C}$ is related to the number of array elements, array arrangement, and array aperture. In an ideal situation, we would like the weight matrix $\bm{C}$ to be the identity matrix $\bm{E}$, as it can highlight the current signal direction without introducing signals from other directions. Therefore, we evaluate the spatial signal focusing ability (SSFA) of the weight matrix $\bm{C}$ by calculating the following formula:
\begin{equation}
    SSFA=|\bm{BA}-\bm{E}|_F
\end{equation}

\begin{figure}[t] 
\centering 
\includegraphics[width=0.48\textwidth]{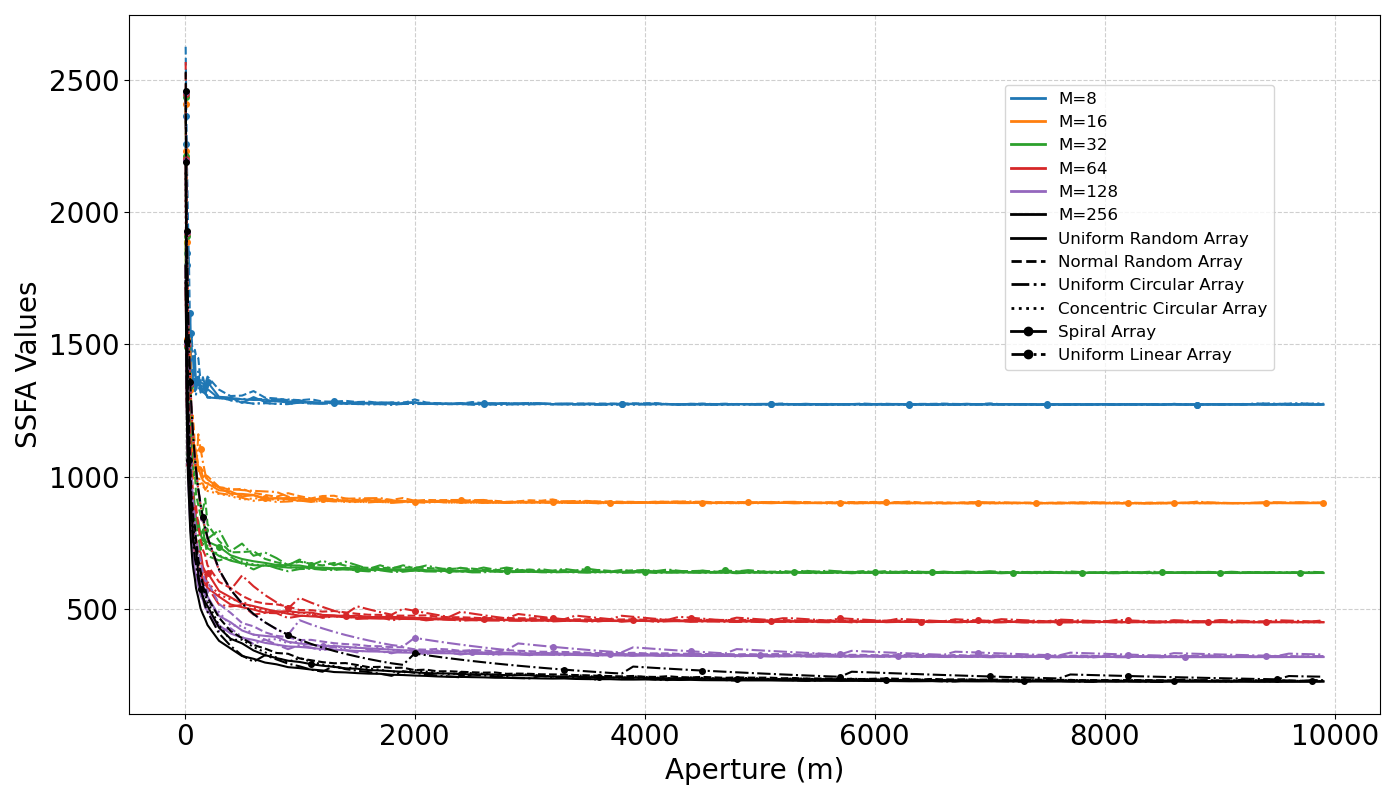} 
\caption{Effects of array on spatial signal focusing.}  
\label{fig:effects_of_array_on_spatial_signal_focusing} 
\end{figure} 

Figure~\ref{fig:effects_of_array_on_spatial_signal_focusing} shows the effect of array on spatial signal focusing under different numbers of elements, array arrangements, and array apertures. In Figure~\ref{fig:effects_of_array_on_spatial_signal_focusing}, we observe that the arrangement of the array has little effect on the spatial signal focusing. In addition, as the array aperture increases, the SSFA first rapidly decreases and then reaches a certain value before remaining stable. The number of array elements has a significant impact on the focusing of spatial signals. The larger the number of array elements, the smaller the SSFA value, and the smaller the mutual interference between signals.

\subsection{Effects of array on noise suppression}
The suppression of noise is another important aspect in the proposed algorithm, which is related to the spatial filter $\bm{B}$. The solution of the spatial filter $\bm{B}$ is related to the number of elements, array arrangement, and array aperture. In an ideal situation, we hope that noise can be eliminated, that is, $\bm{BN}=\bm{0}$, so we evaluate the noise suppression ability (NSA) of the spatial filter $\bm{B}$ by the following formula:
\begin{equation}
    NSA = \frac{\sum_{i=1}^T |\bm{BN_i}-\bm{0}|_F}{T},
\end{equation}
where $\bm{N}_i$ represents the $i$-th snapshot in $\bm{N}$.

\begin{figure}[t] 
\centering 
\includegraphics[width=0.48\textwidth]{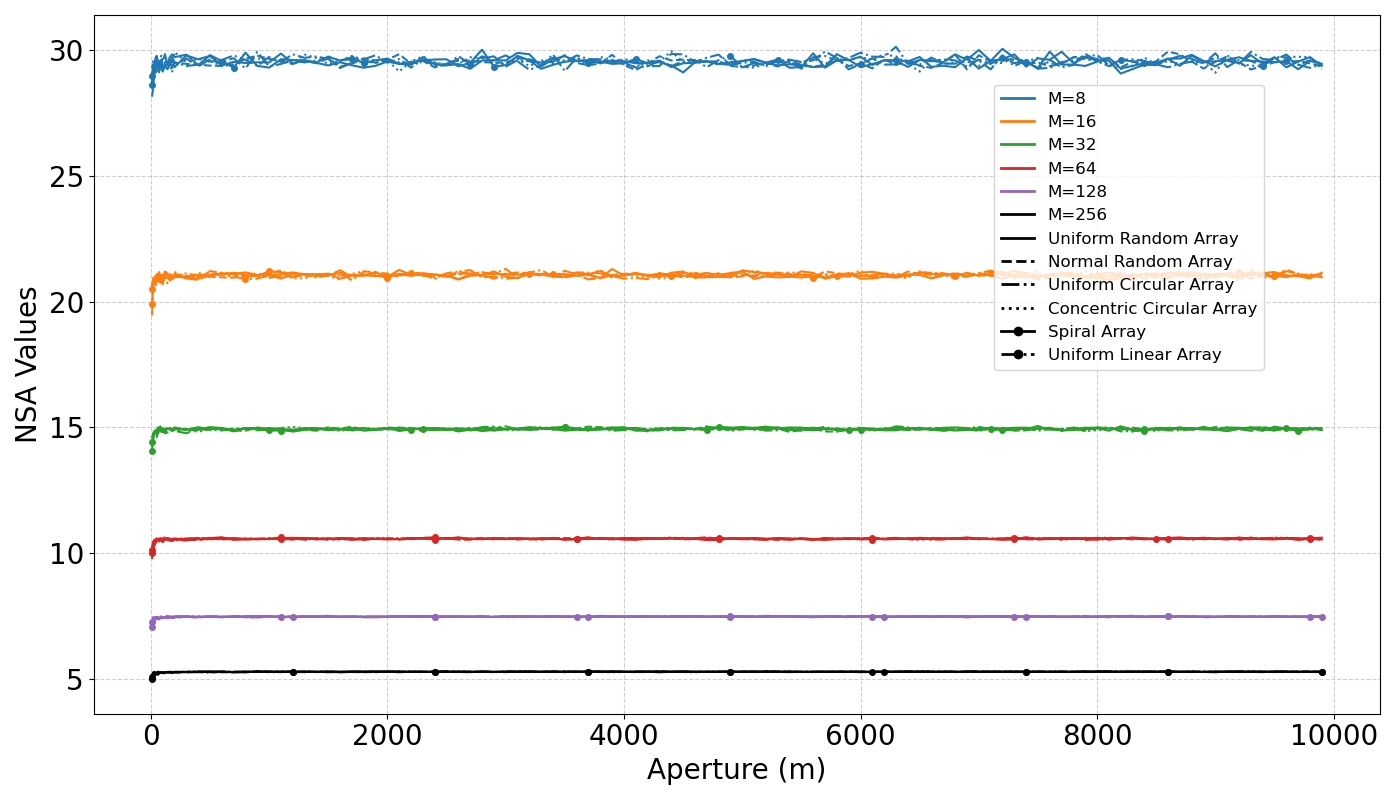} 
\caption{Effects of array on noise suppression.}  
\label{fig:effects_of_array_on_noise_suppression} 
\end{figure} 

We assume that the noise is white noise. Figure~\ref{fig:effects_of_array_on_noise_suppression} shows the impact of the array on noise suppression under different numbers of elements, array arrangements, and array apertures. In Figure~\ref{fig:effects_of_array_on_noise_suppression}, we observe that the arrangement of the array and the aperture of the array have a relatively small impact on noise suppression, while the number of elements in the array has a greater impact on noise suppression. The more elements there are, the better the ability to suppress noise.

\section{Experiments for Algorithm Performance}
In this section, we mainly explore the influence of different signal numbers and iteration times on the proposed algorithm through simulation experiments, and compare the proposed algorithm with MUSIC, CBF, MVDR, and L1SVD algorithms under different signal-to-noise ratios, different snapshot numbers, and signal coherence conditions.

In the previous section~\ref{subsec:effects of array on q}, we found that the proposed algorithm is applicable to two-dimensional large aperture arrays. Therefore, we assume that the narrowband signal propagation speed $c=1500~m/s$, frequency $f$ is $100~Hz$, and the array aperture and arrangement are two-dimensional uniformly random distribution arrays with an aperture of 8000 meters.

According to Formula~\eqref{eq:BAS=S}, we use the following formula to measure the ability of the proposed algorithm to extract signals ($ESA$, Extracting Signals Ability):
\begin{equation}
    ESA=\frac{\sum_{i=1}^T |\bm{B\widetilde{A}S}_i-\bm{S}_i|_F}{T},
\end{equation}
where $\bm{S}_i$ represents the $i$-th snapshot in $\bm{S}$. At the same time, we also use $NSA$ to measure the proposed algorithm's ability to suppress noise. In addition, according to Formula~\eqref{eq:BX=S}, we use the following formula to comprehensively measure the ability of the proposed algorithm to extract signals and suppress noise in different situations ($CA$, Comprehensive Ability):
\begin{equation}
    CA = \frac{\sum_{i=1}^T |\bm{BX}_i-\bm{S}_i|_F}{T}.
\end{equation}

We use the following formula to represent the average interference energy in each spatial direction after filtering by the spatial filter $\bm{B}$:
\begin{equation}
    \overline{I} = \frac{\sum|\bm{BAS}-\bm{S}|^2}{TR},
\end{equation}
where the symbol $\sum$ represents summing up each element in the matrix. Use the following formula to represent the average noise energy in each spatial direction after filtering through a spatial filter $\bm{B}$:
\begin{equation}
    \overline{N} = \frac{\sum|\bm{BN}|^2}{TR}.
\end{equation}
For the average signal energy with signal direction in space, we use the following formula:
\begin{equation}
    \overline{S} = \frac{\sum|\bm{S}|^2}{TK}
\end{equation}

$ESA$, $NSA$, and $CA$ evaluated the ability of spatial filters to suppress noise and eliminate mutual interference between signals. To evaluate the ability of spatial filters to distinguish signals from interference and noise in space, we use the following formulas to represent the ratio of signal energy to noise energy ($SNR_B$), the ratio of signal energy to interference energy ($SIR_B$), and the ratio of signal energy to noise plus interference energy ($SNIR_B$) averaged in each spatial direction after filtering by a spatial filter $\bm{B}$:
\begin{equation}
    SIR_B = 10\log_{10}\frac{\overline{S}}{\overline{I}},
\end{equation}

\begin{equation}
    SNR_B = 10\log_{10}\frac{\overline{S}}{\overline{N}},
\end{equation}
and
\begin{equation}
    SNIR_B = 10\log_{10}\frac{\overline{S}}{\overline{I}+\overline{N}}.
\end{equation}
$SIR_B$, $SNR_B$, and $SNIR_B$ respectively represent the ability of spatial filter $\bm{B}$ to distinguish signals from noise and interference in space.

In large aperture arrays, the resolution of the array is relatively high; therefore, when comparing with other algorithms, we use the following formula to measure the performance of different algorithms in DOA estimation (COR, the Ratio of average energy in the Correct direction to average energy in Other directions):
\begin{equation}
\label{eq:COR}
    COR=10 \log_{10}{\frac{(R-K)\sum_{i \in \theta_s}P_i}{K\sum_{i \notin \theta_s}P_i}},
\end{equation}
where $P$ represents the spatial energy spectrum obtained by each algorithm, and $P_i$ represents the energy value in the $i$-th direction of the spatial energy spectrum $P$. For the algorithm we proposed, there is:
\begin{equation}
    P = \overline{|\bm{BX}|^2}.
\end{equation}

The meaning of Formula~\eqref{eq:COR} is the ratio of the average energy in the correct signal direction to the average energy in other directions in the spatial energy spectrum. Considering that the signal is sparse in the spatial direction, Formula~\eqref{eq:COR} simultaneously measures the accuracy and resolution of the algorithm in estimating DOA.

\begin{figure}[t] 
\centering 
\includegraphics[width=0.48\textwidth]{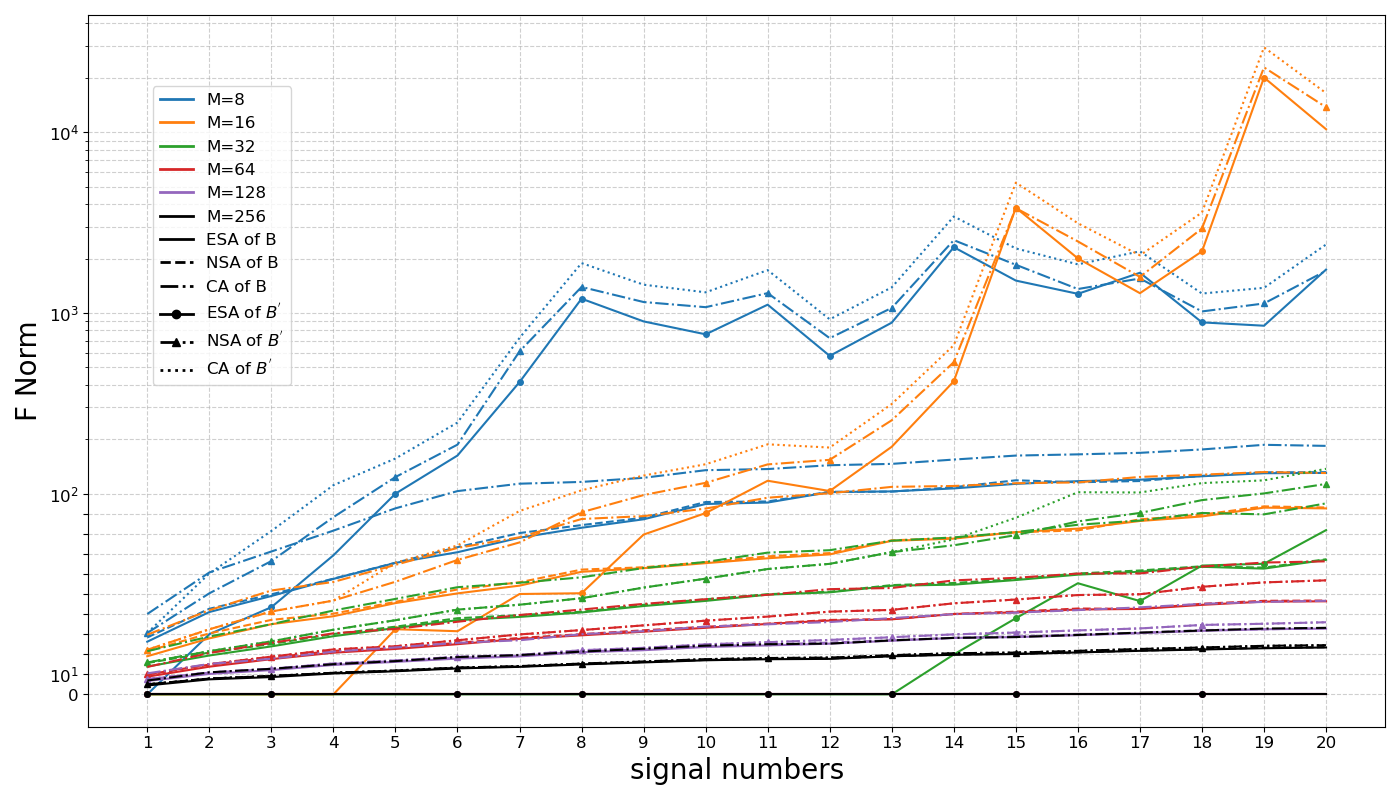} 
\caption{Explore the influence of the number of signals on the proposed algorithm on $ESA$, $NSA$ and $CA$.}  
\label{fig:target_number} 
\end{figure}

\begin{figure}[t] 
\centering 
\includegraphics[width=0.48\textwidth]{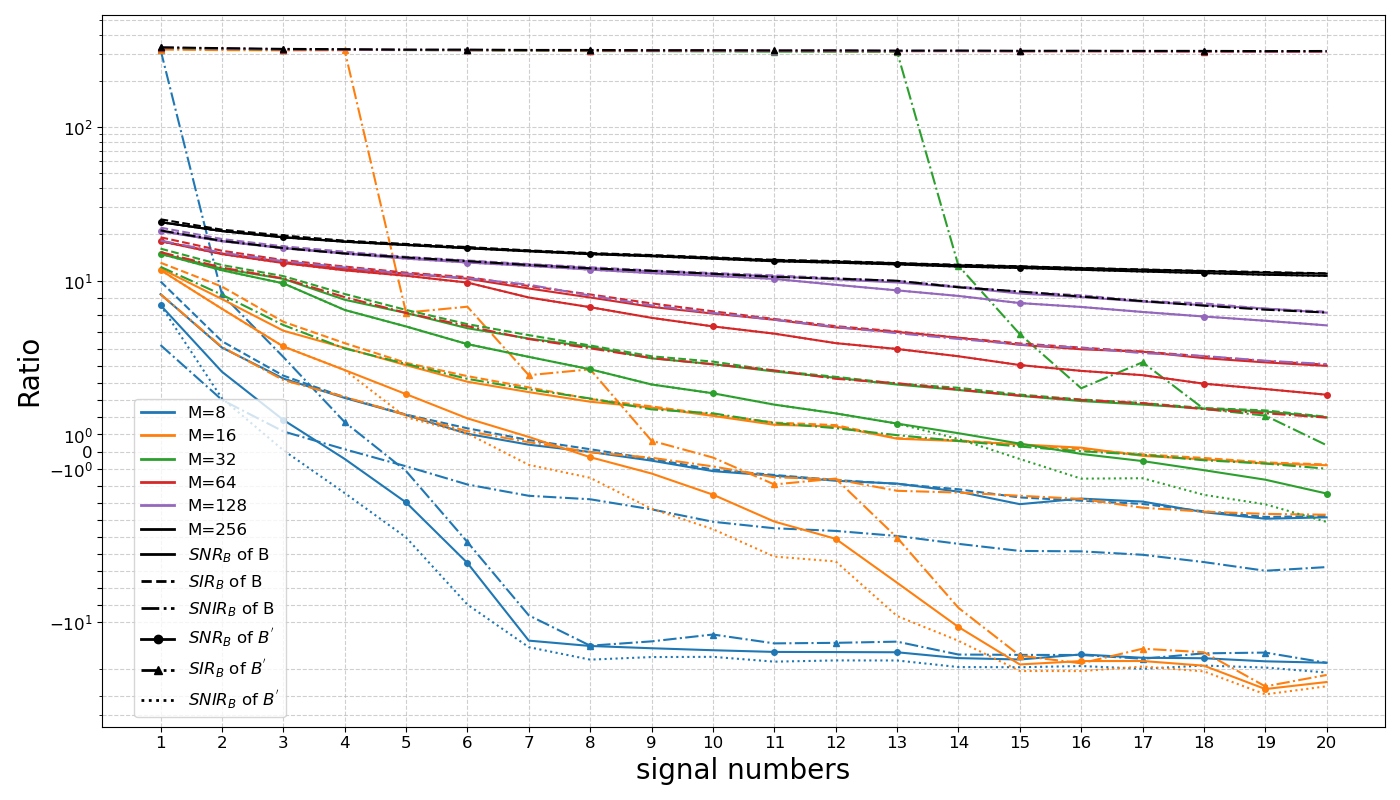} 
\caption{Explore the influence of the number of signals on the proposed algorithm on $SNR_B$, $SIR_B$ and $SNIR_B$.}  
\label{fig:target_number_ratio} 
\end{figure} 

\begin{figure}[t] 
\centering 
\includegraphics[width=0.48\textwidth]{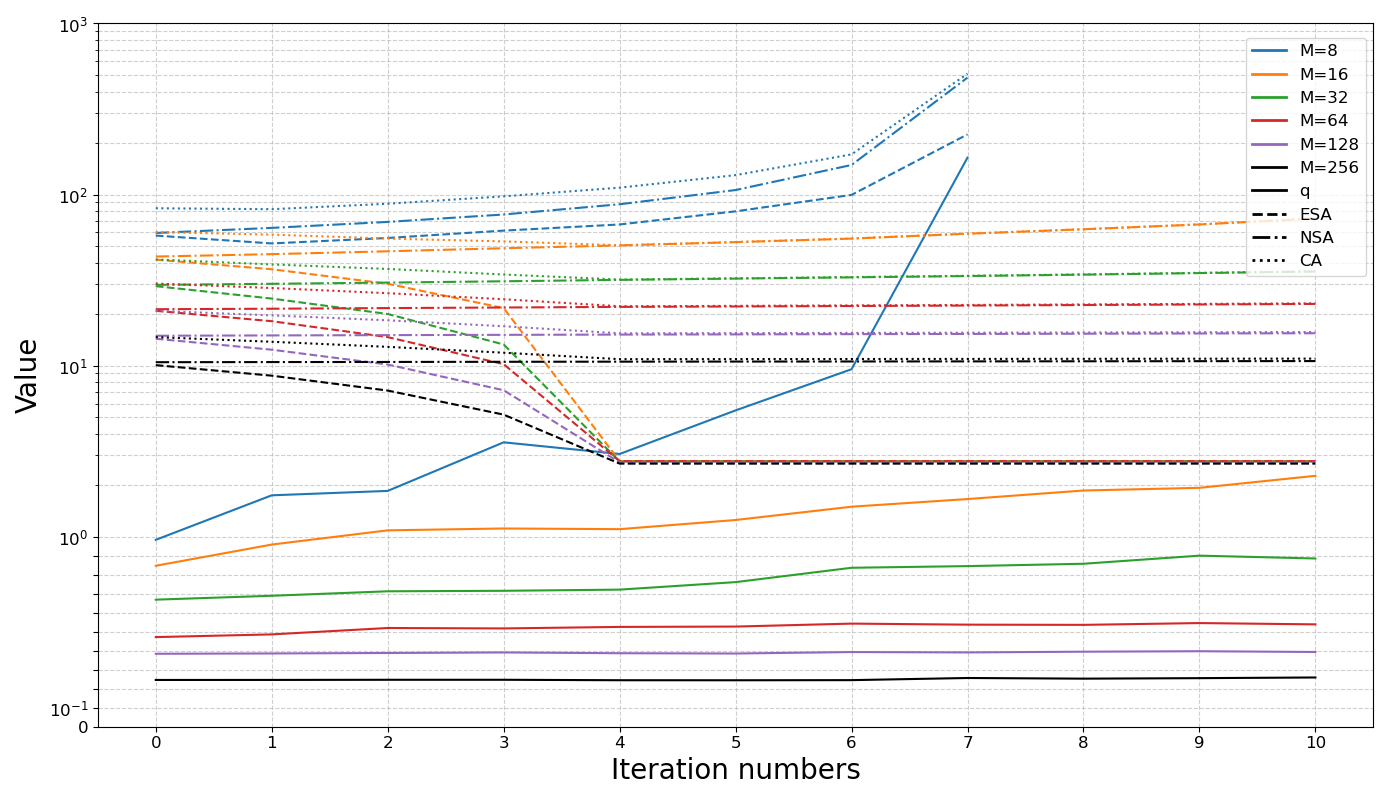} 
\caption{Explore the influence of the number of iteration on the proposed algorithm.}  
\label{fig:iteration_count} 
\end{figure} 

\subsection{Number of signals}
In section~\ref{sec:Effects of Array on Algorithm Performance}, we studied the impact of arrays on algorithms, and in this section, we explore the influence of the signals number on the proposed algorithm. We scale the number of signals from 1 to 20 and use $ESA$, $NSA$, and $CA$ metrics to evaluate the performance of the algorithm. During the experiment, we set the signal-to-noise ratio to 0 dB. As shown in Figure~\ref{fig:target_number}, we demonstrate the variation of the $ESA$, $NSA$ and $CA$ indices with respect to the number of signals for the spatial filter $\bm{B}$ obtained by optimizing Formula~\eqref{optim:min |b_i1|} before the iteration begins, which satisfies Formula~\eqref{eq:BAS&BN approx 0}, and the spatial filter $\bm{B^{'}}$ obtained by optimizing Formula~\eqref{optim:min b_i1^2+...+b_im^2_final} after an appropriate number of iterations (min (K, M-1)), which satisfies Formula~\eqref{eq:BN approx 0}.

In Figure~\ref{fig:target_number}, as the number of signals increases, the $NSA$ values of spatial filters $\bm{B}$ and $\bm{B}^{'}$ become larger. This is because we fix the signal-to-noise ratio at 0 dB, so as the number of signals increases, the noise also gradually increases. In addition, as the number of signals increases, the $ESA$ value corresponding to spatial filter $\bm{B}$ gradually increases, while the $ESA$ value corresponding to $\bm{B}^{'}$ first remains unchanged and then increases. This is because when the number of signals is not large, the spatial filter $\bm{B}^{'}$ satisfies the core idea that there is a signal in the direction corresponding to the maximum value in the spatial spectrum during the iteration process to eliminate interference between signals. However, when the number of signals is too large and the $q< \frac{1}{K}$ condition is not met, the core idea during the iteration process no longer holds, resulting in an increase in the $ESA$ value.

For indicator $CA$, when the number of signals is not high, spatial filter $\bm{B}^{'}$ performs better than spatial filter $\bm{B}$. When the number of signals increases to a certain value, the performance of spatial filter $\bm{B}$ is better than that of spatial filter $\bm{B}^{'}$. In addition, we also found that as the number of array elements increases, the proposed algorithm can adapt to more signal numbers, and the trend of $CA$ value changes slows down. At a signal-to-noise ratio of 0 dB, arrays with 8, 16, and 32 elements exhibit similar performance for spatial filters $\bm{B}$ and $\bm{B}^{'}$ at signal numbers of 2, 5, and 14, respectively. For arrays with 64, 128, and 256 elements, the spatial filter $\bm{B}^{'}$ obtained iteratively has good performance when the number of signals is less than 20. Figure~\ref{fig:target_number} also shows that the proposed algorithm is more robust when there are multiple array elements.

In Figure~\ref{fig:target_number_ratio}, we use $SIR_B$, $SNR_B$, and $SNIR_B$ metrics to evaluate the performance of the algorithm at a signal-to-noise ratio of 0 dB. Figure~\ref{fig:target_number} and Figure~\ref{fig:target_number_ratio} have similar trends. As the number of signals increases, $SNR_B$, $SIR_B$, and $SNIR_B$ all decrease. The $SIR_B$ of spatial filter $\bm{B^{'}}$ drops sharply from signal number 1 at 8 elements, from 4 at 16 elements, and from 13 at 32 elements. The $SNIR_B$ of spatial filter $\bm{B^{'}}$ is less than 0 starting from signal number 3 at 8 elements, from 6 at 16 elements, and from 14 at 32 elements. The smaller the $SNIR_B$, the more difficult it is to distinguish signals from noise and interference. In the case of an ideal number of signals, the proposed algorithm can eliminate interference between signals. Furthermore, although the signal-to-noise ratio was set to 0 dB in the experiment, the $SNR_B$ after filtering by spatial filter $\bm{B^{'}}$ can be greater than 0 dB due to the ability of spatial filter to suppress noise.

\subsection{Iteration count}

In this section, we will explore the impact of iteration count on the algorithm. We assume that the number of signals is 4 and scale the number of iterations from 1 to 10. The iteration count of 0 represents the preliminary estimation of spatial filter $\bm{B}$ based on optimization Formula~\eqref{optim:min |b_i1|}, which conforms to Formula~\eqref{eq:BAS&BN approx 0}. During the experiment, we set the signal-to-noise ratio to 0 dB. As shown in Figure~\ref{fig:iteration_count}, we demonstrate the impact of iteration times on $q$, $ESA$, $NSA$, and $CA$ metrics. Because the spatial filter $\bm{B}$ is updated during the iteration process, the value of $q$ also changes with the update of the spatial filter. It is worth noting that every time $B$ is updated to block signals in one direction, the number of signals $K$ participating in the calculation of $q<\frac{1}{K}$ condition will also decrease by one, because the blocked signals will not interfere with signals in other directions.

In Figure~\ref{fig:iteration_count}, we observe that for a signal number of 4, an 8-element array is not suitable (Figure~\ref{fig:target_number} shows that an 8-element array can accommodate up to two signal numbers at most), so the error increases with each iteration starting from 1 in the 8-element array. For 16-256 array elements, as the number of iterations increases, $ESA$ becomes smaller until the final signal is found in the fourth iteration, after which the excess number of iterations has almost no effect on $ESA$. As the number of iterations increases, $NSA$ and $CA$ gradually increase, mainly reflected in the gradual weakening of noise suppression. For arrays with a large number of elements, the excess number of iterations has less impact on $NSA$ and $CA$. The more elements there are, the smaller the impact. In addition, as the iteration progresses, the value of $q$ also increases, but the number of signals $K$ involved in the calculation also decreases. The more elements there are in the array, the slower the change in the value of $q$, and the less impact it has on the algorithm.

It is worth noting that $q<\frac{1}{K}$ is a sufficient condition for the presence of signals in the direction corresponding to the maximum value of the spatial spectrum. Even if $q<\frac{1}{K}$ does not hold, there may still be signals in the direction corresponding to the maximum value of the spatial spectrum. As shown in the curve of $q$ with 16 elements in Figure~\ref{fig:iteration_count}, even if $q>1$ after the second iteration, the proposed algorithm still holds true.

By summarizing the previous discussions, we found that the proposed method has better robustness and performance in multi-element large aperture two-dimensional arrays.

\subsection{Time complexity}

\begin{table}
\caption{The time complexity of the algorithms} 
\centering
\label{tabel:time_complexity}
\begin{tabular}
{|p{0.25\linewidth}|p{0.55\linewidth}|}
\hline
Algorithms & Time complexity \\ \hline
Proposed      & $O(M^2(T+R)I)$    \\ \hline
MVDR      & $O(M^3+M^2(T+R))$ \\ \hline
MUSIC      & $O(M^3+M^2(T+R))$ \\ \hline
CBF      & $O(M^2(T+R))$ \\ \hline
L1SVD      & $O(R^{3.5})$ \\ \hline
\end{tabular}
\end{table}

Time complexity is an important indicator for measuring algorithm performance. As shown in Table 1, we compared the proposed algorithm with CBF, MVDR, MUSIC, and L1SVD algorithms in terms of time complexity. From the table, it can be found that CBF has the lowest time complexity, L1SVD has the highest time complexity, and the proposed algorithm has a moderate time complexity that is related to the number of iterations. The time complexity of the proposed algorithm is on a similar scale to MVDR and MUSIC algorithms.

\subsection{Snapshot counts}

\begin{figure}[t] 
\centering 
\includegraphics[width=0.48\textwidth]{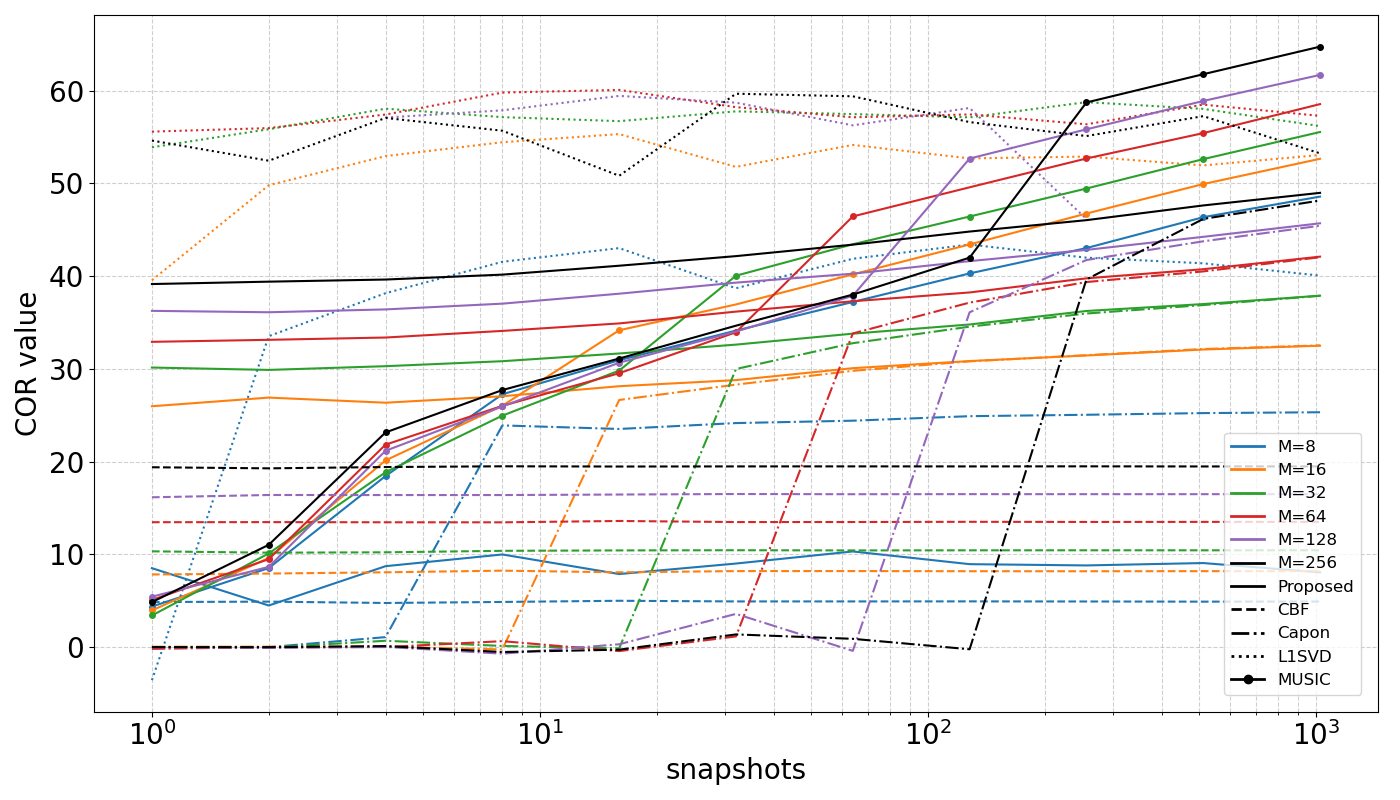} 
\caption{Explore the influence of the snapshot counts on the proposed algorithm.}  
\label{fig:snapshots} 
\end{figure} 

\begin{figure}[t] 
\centering 
\includegraphics[width=0.48\textwidth]{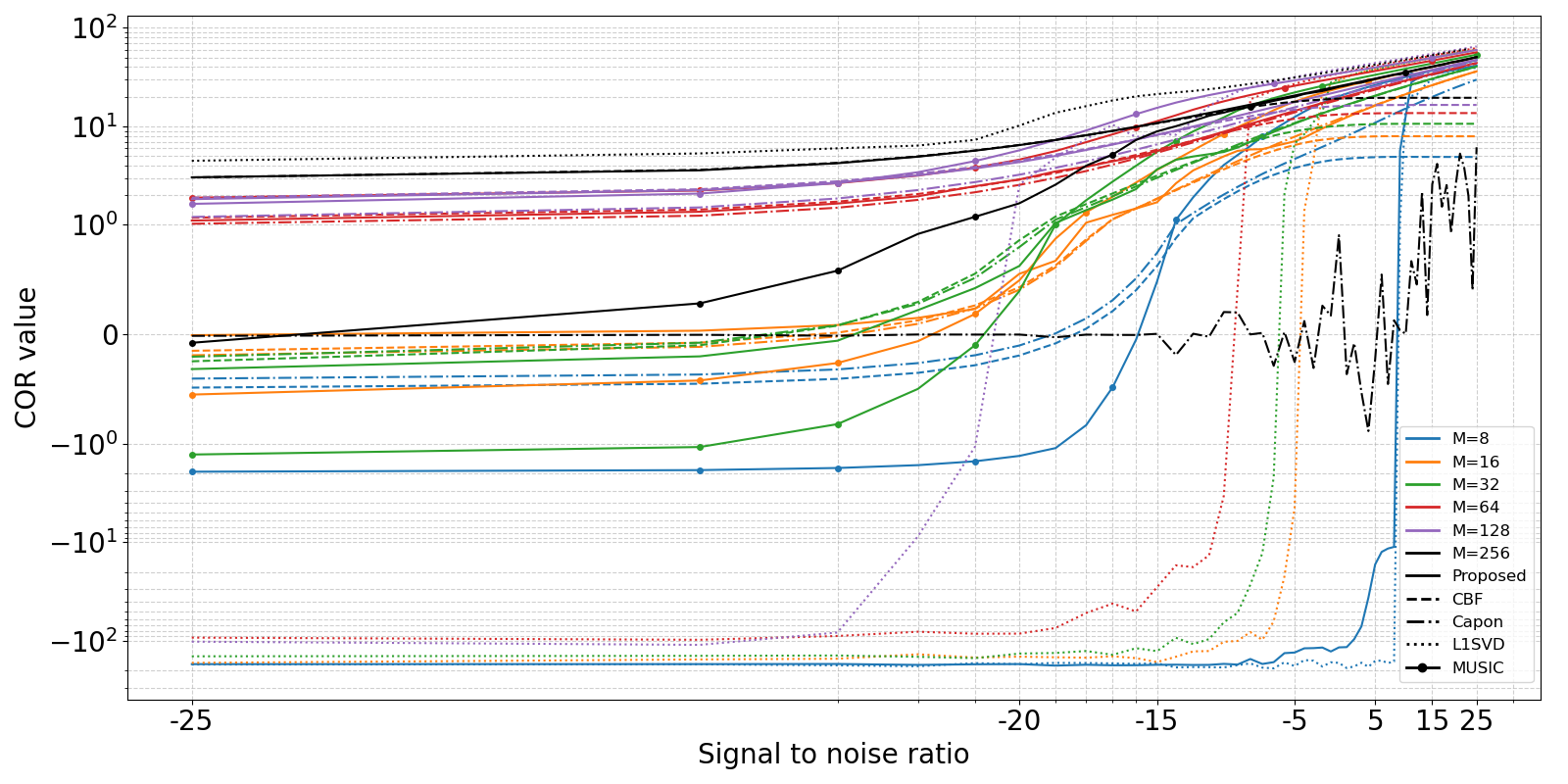} 
\caption{Explore the influence of the Signal-to-noise ratio on the proposed algorithm.}  
\label{fig:Signal-to-noise ratio} 
\end{figure} 

\begin{figure}[t] 
\centering 
\includegraphics[width=0.48\textwidth]{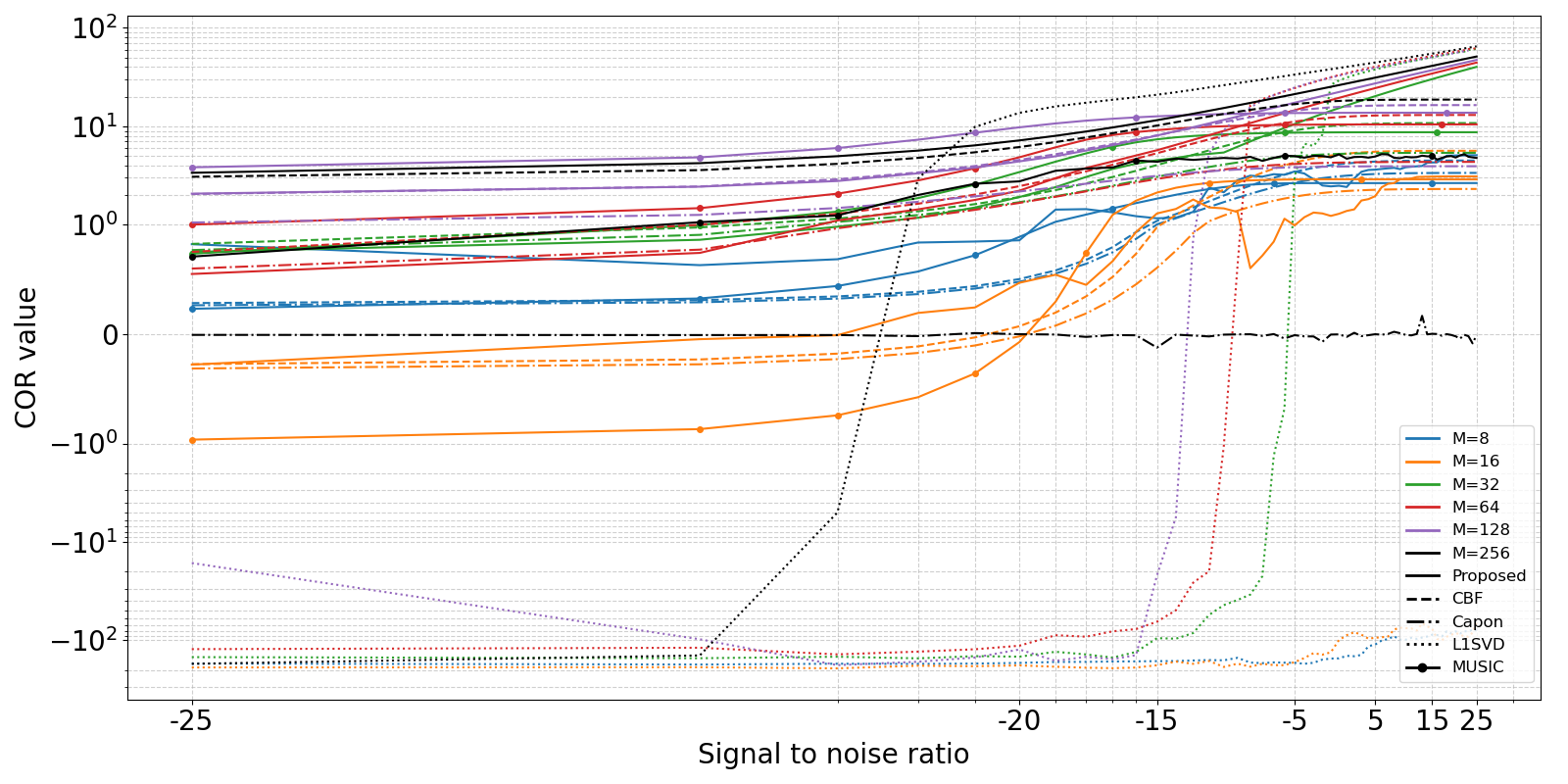} 
\caption{Explore the influence of the coherent signal on the proposed algorithm.}  
\label{fig:coherent_signal} 
\end{figure}

\begin{figure}[t] 
\centering 
\includegraphics[width=0.44\textwidth]{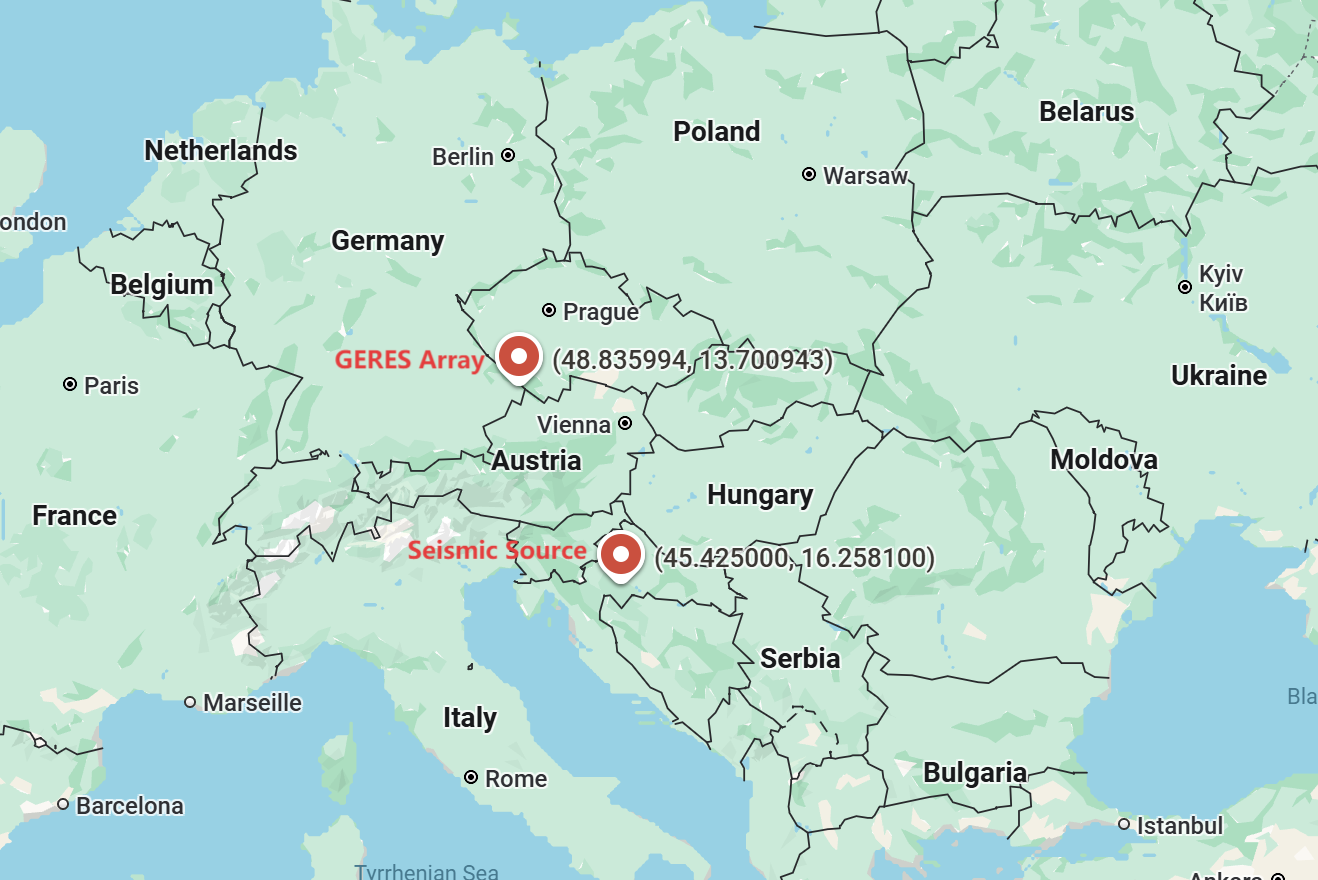} 
\caption{Geographic location of the GERES array and seismic source.}  
\label{fig:map} 
\end{figure}

To investigate the performance of the proposed algorithm under different snapshot counts, we varied the snapshot counts from 1 to 1000. During the experiment, we kept the signals incoherent, set the number of signals to 3, and controlled the signal-to-noise ratio to 20 dB. As shown in Figure~\ref{fig:snapshots}, it compares the performance of different methods under different snapshot counts. From Figure~\ref{fig:snapshots}, we find that in a small number of snapshots, the MVDR algorithm fails due to fewer snapshots than the number of elements, and the MUSIC algorithm performs poorly. However, the proposed algorithm outperforms other algorithms except for L1SVD. Although the L1SVD algorithm performs best with a small number of snapshots, its computational complexity is much greater than the proposed algorithm. Under a large number of snapshots, we observed that as the number of snapshots increased, the performance of the MVDR algorithm approached that of the proposed algorithm, and ultimately, the proposed algorithm and MVDR performed similarly under a large number of snapshots. The L1-SVD and MUSIC algorithms exhibit better performance in a large number of snapshots due to their lower noise content in the spatial spectrum compared to the proposed algorithm and MVDR algorithm. Figure~\ref{fig:snapshots} illustrates that the proposed algorithm has the advantages of low time complexity and good performance in scenes with a small number of snapshots.

\subsection{Signal-to-noise ratio}

To investigate the performance of the proposed algorithm under different signal-to-noise ratios, we varied the signal-to-noise ratio from -25 dB to 25 dB. During the experiment, we kept the signals incoherent and set the number of signals to 3 while controlling the number of snapshots to 200. Figure~\ref{fig:Signal-to-noise ratio} shows a comparison of the performance of different methods under different signal-to-noise ratios. The MUSIC algorithm requires prior knowledge of the number of signals to divide the noise space and signal space. Under low signal-to-noise ratio, the MUSIC algorithm is no longer able to determine the number of signals based on the size of their eigenvalues. The MUSIC algorithm shown in the figure is an artificially given number of signals, and the actual MUSIC algorithm is no longer effective. From Figure~\ref{fig:Signal-to-noise ratio}, it can be seen that the proposed algorithm performs well at low signal-to-noise ratios. In addition, we observed that the proposed algorithm and MVDR algorithm approach each other as the signal-to-noise ratio increases. When the number of elements is less than 256, L1SVD performs the worst at low signal-to-noise ratios. Figure~\ref{fig:Signal-to-noise ratio} illustrates that the proposed algorithm still exhibits good performance even at low signal-to-noise ratios.

\subsection{Coherent signal}

In order to investigate the performance of the proposed algorithm under coherent signals, we kept the coherence between signals in the experiment, set the number of snapshots to 200 and the number of signals to 3. Figure~\ref{fig:coherent_signal} shows a comparison of the performance of different methods under coherent signals and different signal-to-noise ratios. Due to signal coherence, the MUSIC algorithm is no longer able to divide the noise space and signal space based on the size of the eigenvalues during eigenvalue decomposition. Therefore, in practice, the MUSIC algorithm has become ineffective. The MUSIC algorithm in the figure is a manually given number of signals. As shown in Figure~\ref{fig:coherent_signal}, under coherent signals, the proposed method remains effective and performs better than other methods. Figure~\ref{fig:coherent_signal} illustrates that the proposed algorithm is capable of processing coherent signals.

\begin{figure}[t] 
\centering 
\includegraphics[width=0.44\textwidth]{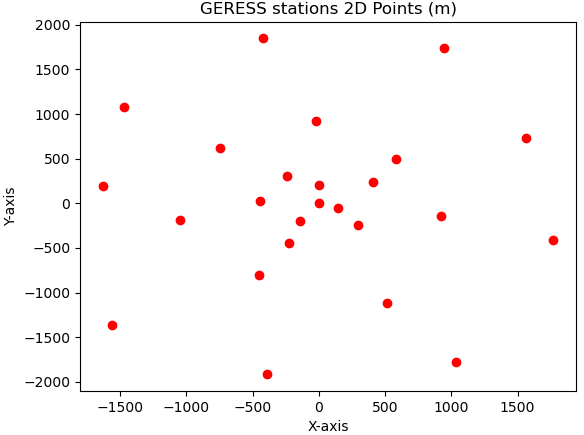} 
\caption{The arrangement of the GERES array.}  
\label{fig:GERESS} 
\end{figure} 

\begin{figure}[t] 
\centering 
\includegraphics[width=0.48\textwidth]{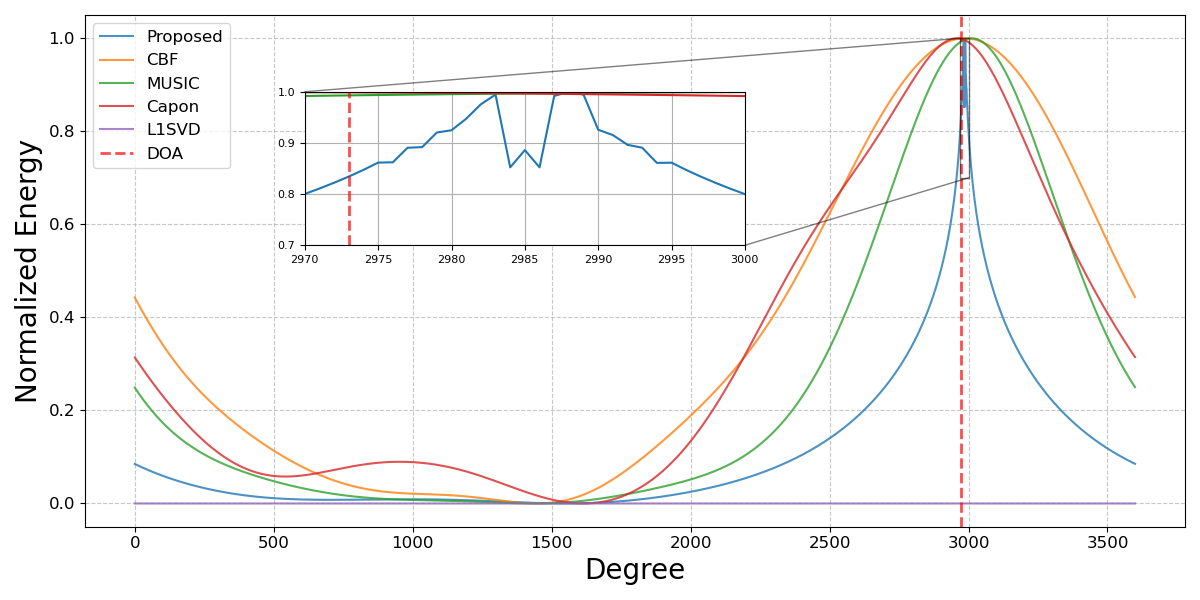} 
\caption{The comparison of different algorithms on seismic wave data.}  
\label{fig:real_data} 
\end{figure}

\begin{figure}[t] 
\centering 
\includegraphics[width=0.38\textwidth]{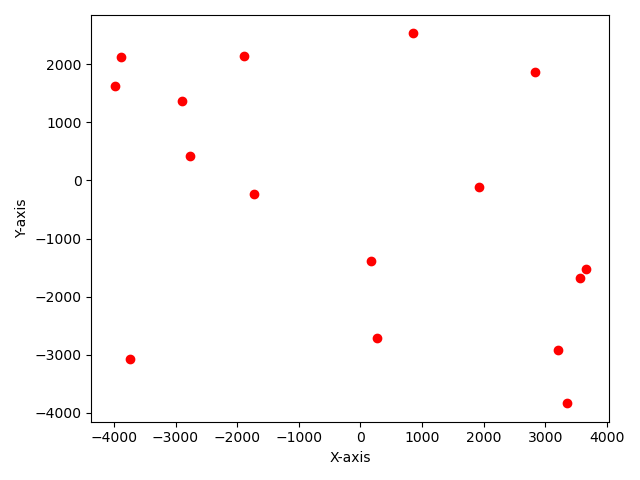} 
\caption{The arrangement of a two-dimensional uniform random distribution array with 16 element.}  
\label{fig:16array} 
\end{figure} 

\begin{figure}[t] 
\centering 
\includegraphics[width=0.45\textwidth]{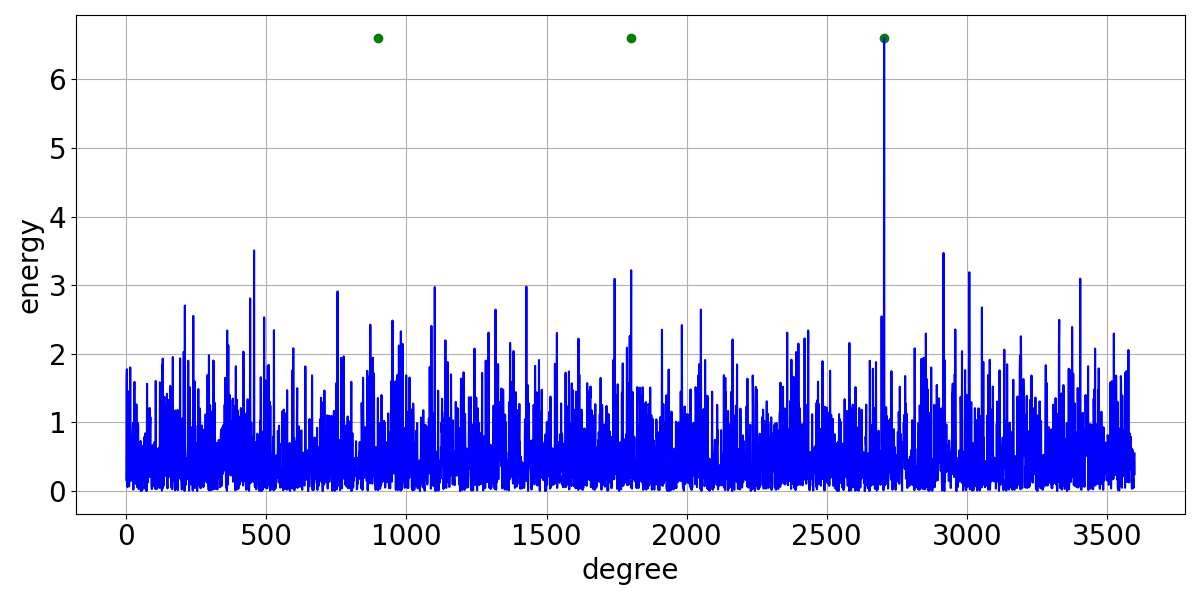} 
\caption{The spatial energy spectrum $\bm{|BX|^2}$ obtained by the spatial filter $\bm{B}$ based on Formula~\eqref{optim:min |b_i1|}. $\theta_0=\emptyset$. $i^*=2705$.}  
\label{fig:initial} 
\end{figure} 

\begin{figure}[t] 
\centering 
\includegraphics[width=0.45\textwidth]{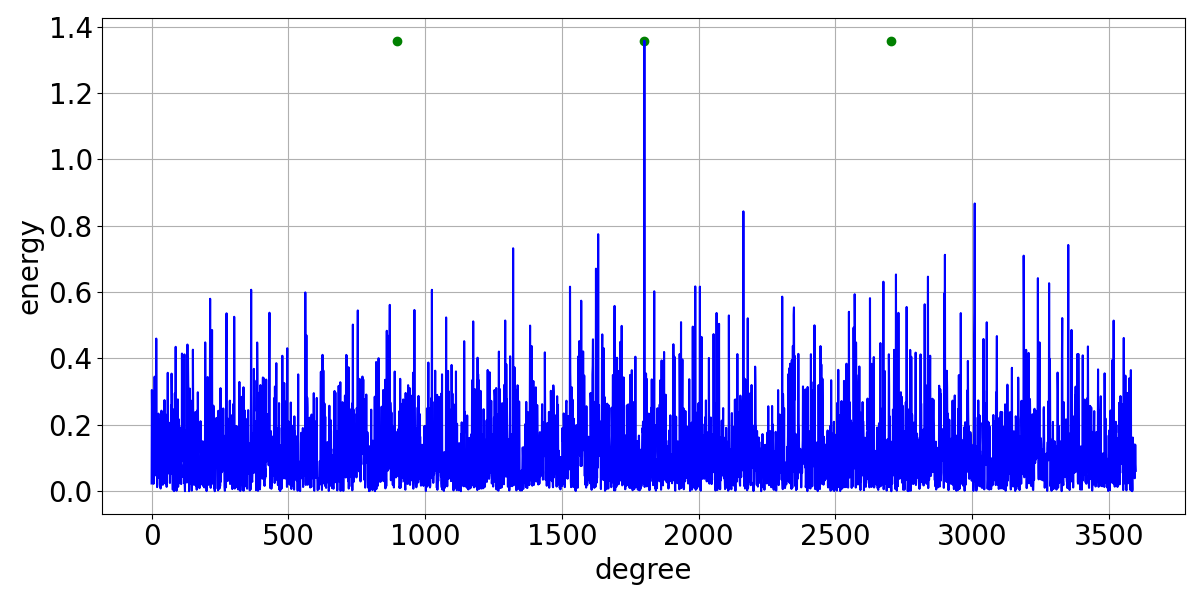} 
\caption{The spatial energy spectrum $\bm{|BX|^2}$ after the first iteration. $\theta_1=\{270.5^{\circ}\}$. $i^*=1802$.}  
\label{fig:iteration1} 
\end{figure} 

\begin{figure}[t] 
\centering 
\includegraphics[width=0.45\textwidth]{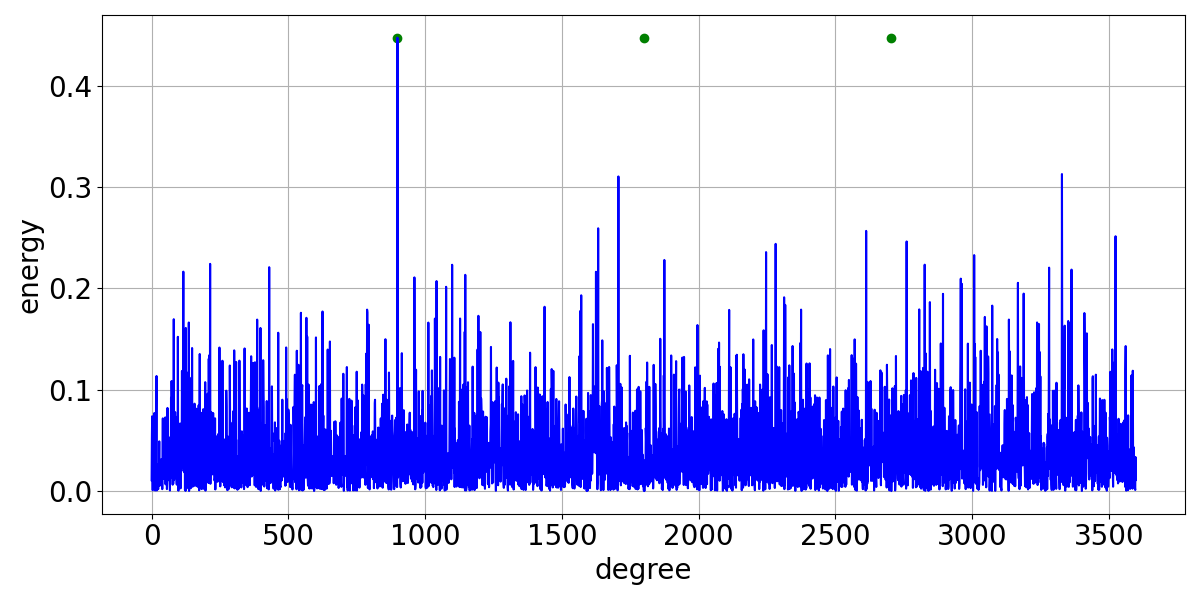} 
\caption{The spatial energy spectrum $\bm{|BX|^2}$ after the second iteration. $\theta_2=\{270.5^{\circ},180.2^{\circ}\}$. $i^*=899$.}  
\label{fig:iteration2} 
\end{figure} 

\subsection{Real data of seismic wave}
To investigate the effectiveness of the proposed algorithm on real data, we used seismic wave data. The seismic wave was generated on December 29, 2020 at 11:19:55 UTC at longitude 42.425 and latitude 16.258, resulting in a magnitude 6.4 earthquake. Figure~\ref{fig:map} shows the geographical location of the seismic wave source.

The seismic wave data were recorded by the German Experimental Seismic System (GERES) array located in the Bavarian Forest, Germany. The array is composed of 25 vertical seismometers. Figure~\ref{fig:map} marks the geographical location of the center of the GERES array with latitude and longitude, while Figure~\ref{fig:GERESS} illustrates the arrangement of the GERES array. Details about the GERES arrays and the exact array configuration can be found in~\cite{harjes1990design}.

During the experiment, we set $f=1Hz$, elevation angle $\varphi = -\frac{\pi}{4}$, and maximum iteration count $I = 3$. For comparison, we normalized the energy spectra of each algorithm, and Figure~\ref{fig:real_data} shows the performance comparison of each algorithm on seismic wave data. We use a red dashed line to represent the DOA of azimuth, which is calculated based on the spherical model using the latitude and longitude of the center of the GERES array and the latitude and longitude of the seismic source. The reason why the proposed algorithm did not form sharp peaks in the energy spectrum is that, according to Formula~\eqref{optim:min b_i1^2+...+b_im^2_final}, the algorithm proposed for finding spatial filter $\bm{B}$ shields energy in some directions, resulting in lower energy in the shielded directions when forming the spatial spectrum. And this also makes the energy spectrum of the proposed algorithm narrower overall compared to other methods. As shown in Figure~\ref{fig:real_data}, the L1SVD algorithm is ineffective and the proposed algorithm demonstrated good performance.

\section{Conclusion}
In this article, we construct the concept of the optimal spatial filter to solve the DOA estimation problem by utilizing the sparsity of spatial signals. By utilizing the concept of the optimal spatial filter, we have transformed the DOA estimation problem into a solution problem for the optimal spatial filter. We discussed the existence of the optimal filter and attempted to solve for it by utilizing spatial signal focusing and noise suppression methods. Due to the unknowability of noise, although we cannot directly solve for the optimal spatial filter, we have provided an approximate solution for the optimal spatial filter by iteratively optimizing the preliminary obtained spatial filter. Through experiments, it was found that the proposed algorithm is suitable for large aperture two-dimensional arrays and has better performance than other algorithms under conditions of fewer snapshots, low signal-to-noise ratio, and coherent signals. In addition, our proposed algorithm does not require prior knowledge of the number of signals and has a relatively low computational complexity. Our contribution lies in solving the DOA estimation problem from the perspective of the optimal spatial filter and providing a solution method suitable for large aperture two-dimensional arrays. In the future, we will continue to explore better methods for solving the optimal spatial filter.



\section*{Acknowledgments}
This should be a simple paragraph before the References to thank those individuals and institutions who have supported your work on this article.


\appendices
\section{The details of array arrangements}
We will introduce the specific details of the arrangement of two-dimensional uniform random distribution array, two-dimensional normal random distribution array, two-dimensional uniform circular array, two-dimensional uniform concentric circular array, two-dimensional spiral array, and one-dimensional uniform linear array.

\subsection{Two-dimensional uniform random distribution array}
For a given array aperture $V$, the coordinates of each element in the two-dimensional uniform random distribution array satisfy a uniform random distribution, that is, the x and y coordinates are uniformly randomly distributed in $[-\frac{V}{2}, \frac{V}{2}]$.

\subsection{Two-dimensional normal random distribution array}
In a two-dimensional normal random distribution array, we follow the 68-95-99.7 rule of normal distribution to include most of the array elements in $[u-3\sigma, u+3\sigma]$, where $u$ represents the mean of normal distribution and $\sigma$ represents the standard deviation of normal distribution. In our experiment, we use a two-dimensional normal distribution and set $u=0$ and $\sigma=\frac{V}{6}$.

\subsection{Two-dimensional uniform circular array}
For a two-dimensional uniform circular array, we set the radius of the circle to $\frac{V}{2}$ and uniformly arrange each element counterclockwise starting from the element with x and y coordinates $(\frac{V}{2}, 0)$.

\begin{figure}[t] 
\centering 
\includegraphics[width=0.45\textwidth]{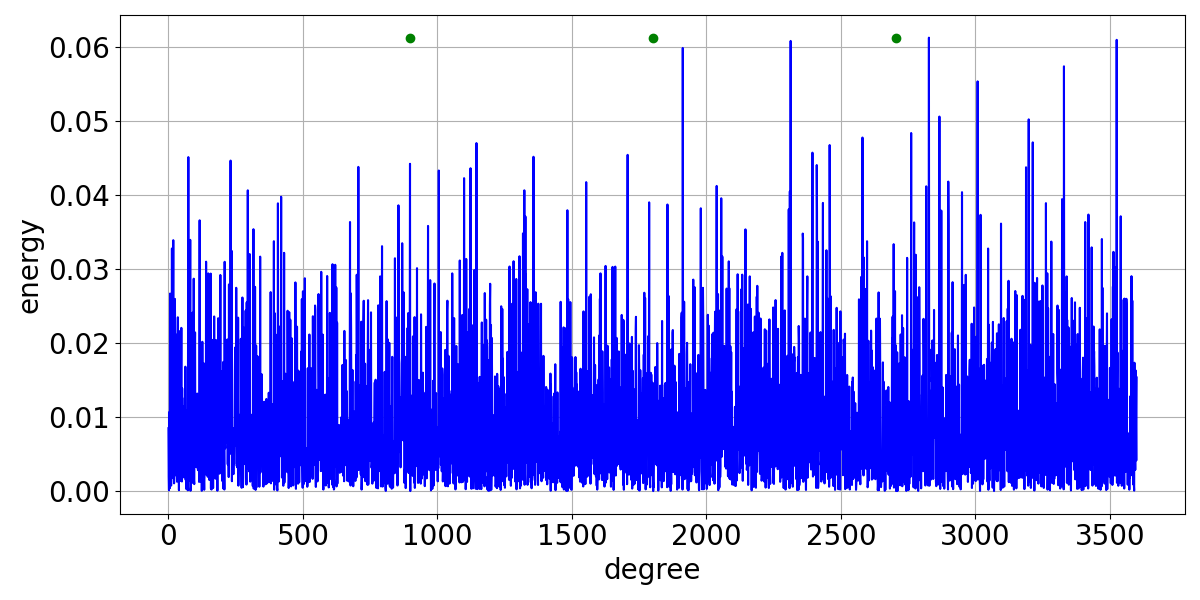} 
\caption{The spatial energy spectrum $\bm{|BX|^2}$ after the third iteration. $\theta_3=\{270.5^{\circ},180.2^{\circ},89.9^{\circ}\}$. $i^*=2827$.}  
\label{fig:iteration3} 
\end{figure} 

\begin{figure}[t] 
\centering 
\includegraphics[width=0.45\textwidth]{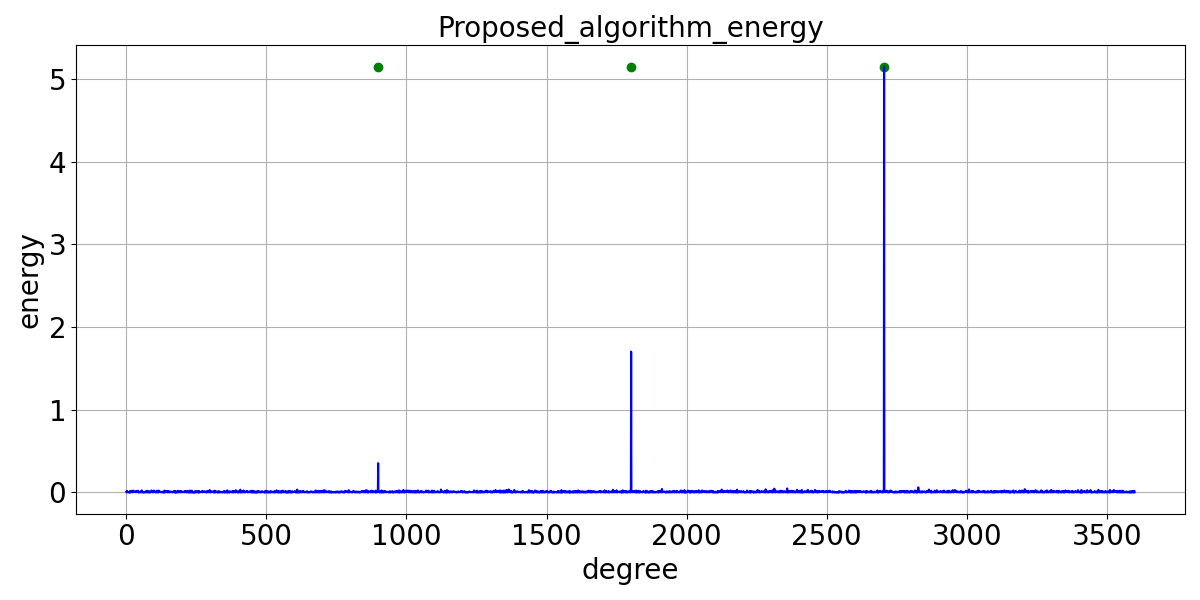} 
\caption{The final spatial energy spectrum $|\bm{BX}|^2$ of proposed algorithm.}  
\label{fig:16proposed_energy} 
\end{figure}

\begin{figure}[t] 
\centering 
\includegraphics[width=0.45\textwidth]{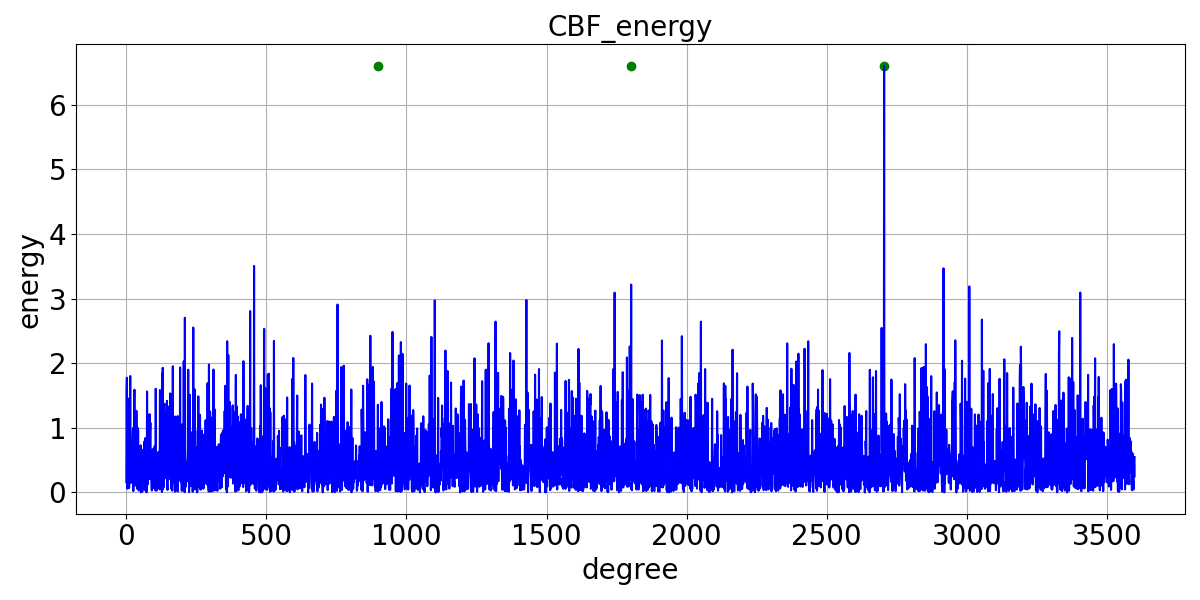} 
\caption{The spatial energy spectrum of CBF algorithm.}  
\label{fig:16CBF_energy} 
\end{figure}

\begin{figure}[t] 
\centering 
\includegraphics[width=0.45\textwidth]{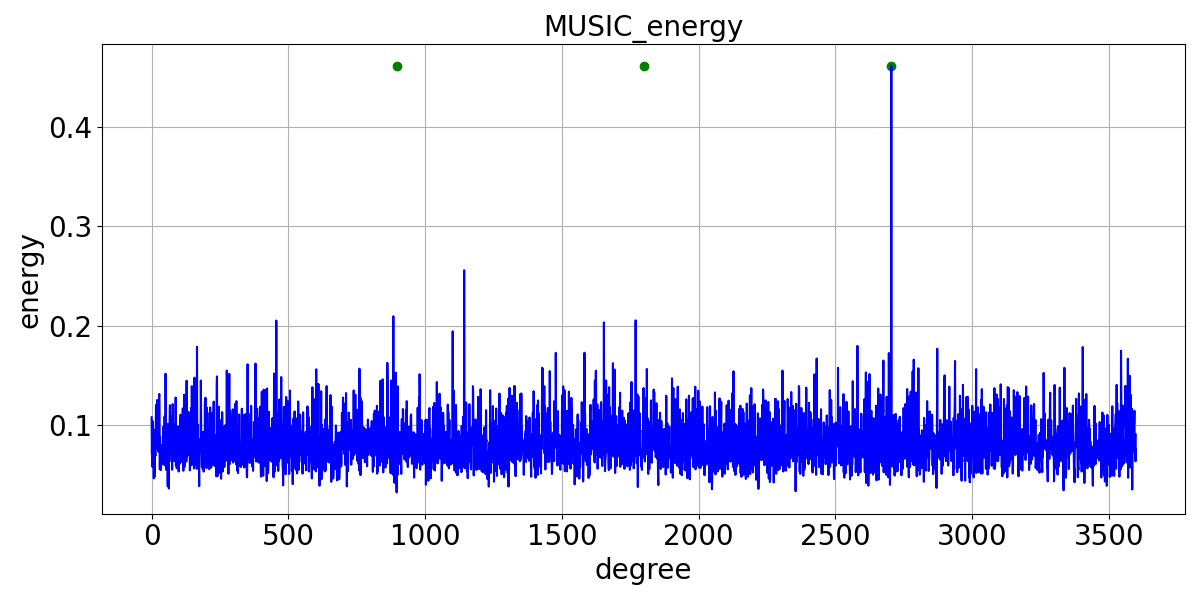} 
\caption{The spatial energy spectrum of MUSIC algorithm.}  
\label{fig:16MUSIC_energy} 
\end{figure}

\begin{figure}[t] 
\centering 
\includegraphics[width=0.45\textwidth]{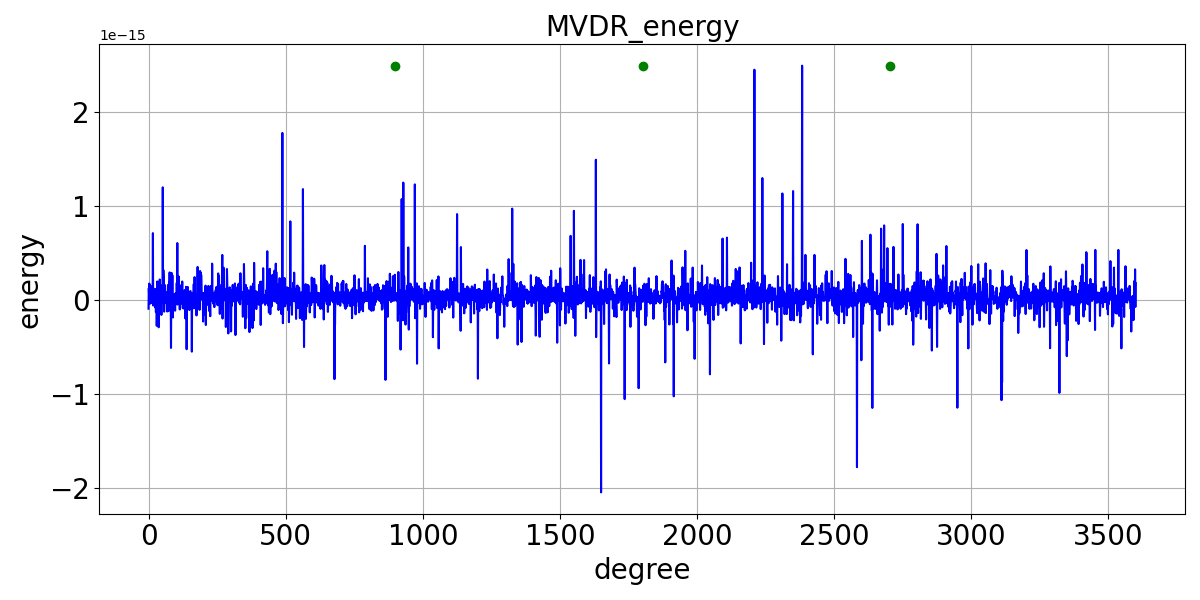} 
\caption{The spatial energy spectrum of MVDR algorithm.}  
\label{fig:16MVDR_energy} 
\end{figure} 

\begin{figure}[t] 
\centering 
\includegraphics[width=0.45\textwidth]{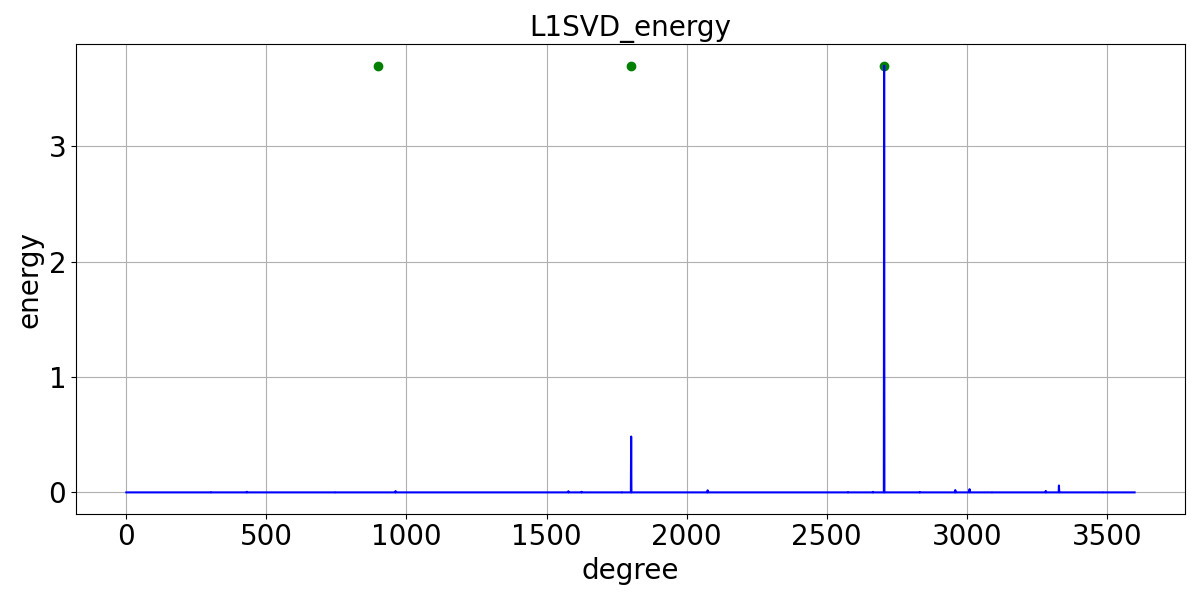} 
\caption{The spatial energy spectrum of L1SVD algorithm.}  
\label{fig:16L1SVD_energy} 
\end{figure} 

\subsection{Two-dimensional uniform concentric circular array}
A two-dimensional uniform concentric circular array is composed of multiple nested two-dimensional uniform circular arrays that share a common center. Each uniform circular array consists of 8 elements, with a total of $\frac{M}{8}$ concentric circular arrays. Each layer of circular array is evenly spaced with a distance of $\frac{4V}{M}$. The radius of the innermost circular array is $\frac{4V}{M}$, and the radius of the outermost circular array is $\frac{V}{2}$. The y-coordinate of the starting element of each layer of circular array is 0, and the x-coordinate is the radius of that layer of circular array.

\subsection{Two-dimensional spiral array}
The two-dimensional spiral array consists of 8 spiral lines, each with the same number of elements in the field. In polar coordinates, the angle $\theta$ of each spiral increases linearly with the radius $r$. For the $k$-th spiral line, its angle consists of two parts, namely the initial angle $\theta_k=k \times \frac{2\pi}{8}$ and the radius dependent angle increment $c \times r$ ($c$ controls the tightness of the spiral, and $r$ is the current radius). Therefore, the polar coordinate equation for the angle $\theta$ at any point on the $k$-th branch is:
\begin{equation}
\label{eq:theta(r,k)}
    \theta(r,k)=\theta_k+c \times r = \frac{2\pi k}{8} + c \times r
\end{equation}

In our experiment, $c$ was set to 0.5, and the array elements on each branch were uniformly distributed with a radius $r$ in polar coordinates. The radius of the outermost array element on each branch in polar coordinates was $r = \frac{V}{2}$, and the angle of each array element was calculated using Formula \eqref{eq:theta(r,k)}.

\subsection{One-dimensional uniform linear array}
For a two-dimensional uniform circular array, the elements in this array are evenly spaced on a line segment with a length of $V$. The center point of the line segment is at the origin of the coordinates.

\section{An example of the proposed algorithm}
We will introduce an example of the proposed algorithm under the simulation experiment. During the experiment, we set the frequency f to 100 Hz, propagation speed to 1500 m/s, number of snapshots to 1, signal-to-noise ratio to 20 dB, $\delta$ to $0.1^{\circ}$, and number of signal sources to 3. We use a two-dimensional uniform random distribution array with 16 elements, and the specific arrangement of the elements is shown in Figure~\ref{fig:16array}. The directions of the signal in space are $89.9^{\circ}$, $180.2^{\circ}$, and $270.5^{\circ}$, respectively.

In Figure~\ref{fig:initial}, we show the spatial energy spectrum $\bm{|BX|^2}$ obtained by the spatial filter $\bm{B}$ based on Formula~\eqref{optim:min |b_i1|}. The green dots in the figure represent the directional positions of the three signals in space. Due to the weak spatial aggregation ability of the few element array, there is mutual interference of spatial signals in Figure~\ref{fig:initial}. According to our proposed method, assuming $q<\frac{1}{K}$ holds under large aperture two-dimensional array, there is a signal in the spatial direction corresponding to the maximum value of the spatial spectrum. Therefore, we iteratively optimize the spatial filter $\bm{B}$. At this point, according to the direction corresponding to the maximum value in Figure~\ref{fig:initial}, $i^*=2705$, $\theta_1=\{270.5^{\circ}\}$.

Then we iterate the spatial filter $\bm{B}$ according to Algorithm~\ref{alg:spatial signal focusing and noise suppression}. Figure~\ref{fig:iteration1},~\ref{fig:iteration2} and~\ref{fig:iteration3} show the spatial energy spectra $\bm{|BX|^2}$ of spatial filter $\bm{B}$ obtained according to Formula~\eqref{optim:min b_i1^2+...+b_im^2} after 1-3 iterations. It is worth noting that as the number of iterations increases, the energy of the spatial spectrum $\bm{|BX|^2}$ decreases. This is because, during the iteration process, the direction of the signal is gradually stripped and shielded. Finally, we obtain $\theta_3=\{270.5^{\circ},180.2^{\circ},89.9^{\circ}\}$. In practical application of the algorithm, because we do not know the number of signal sources, we intentionally iterate multiple times (up to 15 iterations for 16 elements). Here, we intentionally iterate three more times for a total of 6 iterations, and finally obtain $\theta_6=\{270.5^{\circ},180.2^{\circ},89.9^{\circ},282.7^{\circ},235.9^{\circ},231.3^{\circ}\}$.

Finally, we need to find the true direction of the signal theta $\theta_s=\{ 270.5^{\circ},180.2^{\circ},89.9^{\circ}\}$ from the set of candidate signal spatial directions $\theta_6$ that we believe may contain signals. According to Formula~\eqref{optim:min b_i1^2+...+b_im^2_final}, we calculate the final spatial filter $\bm{B}$. Figure~\ref{fig:16proposed_energy} shows the spatial energy spectrum $\bm{|BX|^2}$ of the final spatial filter $\bm{B}$ obtained at a single snapshot count. It can be seen from Figure~\ref{fig:16proposed_energy} that the current snapshot energy in the $89.9^{\circ}$ direction is not very large.

In addition, under the same settings, we also compared the proposed algorithm with other methods. Figure~\ref{fig:16CBF_energy},~\ref{fig:16MUSIC_energy},~\ref{fig:16MVDR_energy} and~\ref{fig:16L1SVD_energy} show the spatial energy spectra of CBF, MUSIC, MVDR, and L1-SVD algorithms, respectively. Under a single snapshot count, the MVDR algorithm fails, the L1SVD algorithm misses signals in the $89.9^{\circ}$ direction, the CBF and MUSIC algorithms perform poorly, and the proposed algorithm performs the best.

\bibliographystyle{IEEEtran}
\bibliography{refs}


\vspace{-22pt}

\begin{IEEEbiography}[{\includegraphics[width=1in,height=1.25in,clip,keepaspectratio]{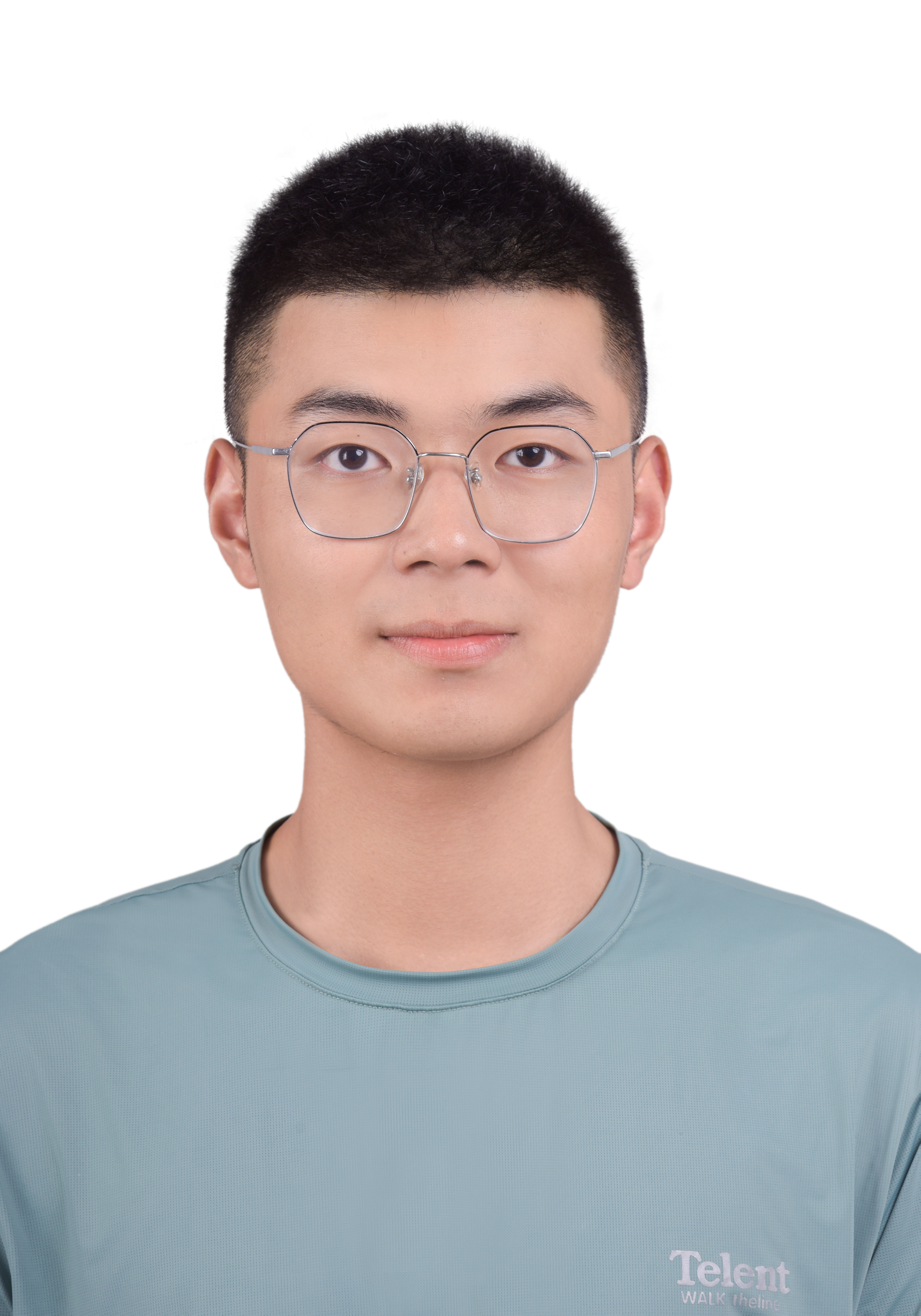}}]{Xuyao Deng} received the bachelor's degree from Shandong University, Shandong, China, in 2023. He is currently a doctoral student at the School of Computer Science, National University of Defense Technology, Changsha, China. His research interests include signal processing, machine learning, and intelligent software systems.

\end{IEEEbiography}

\begin{IEEEbiography}[{\includegraphics[width=1in,height=1.25in,clip,keepaspectratio]{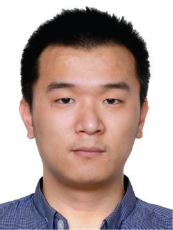}}]{Kele Xu}
(Senior Member, IEEE) received the doctorate degree from Paris VI University, Paris, France, in 2017. He is currently an Associate Professor with the School of Computer Science, National University of Defense Technology, Changsha, China. His research interests include audio signal processing, machine learning, and intelligent software systems. He is the associate editor for IEEE Transactions on Circuits and Systems for Video Technology and Guest editor for Science Partner Journal Cyborg and Bionic Systems. He has (co-)authored more than 100 publications in peer reviewed journals and conference proceedings, including TASLP, TMI, ICML, CVPR, NeuRIPS, ICLR, AAAI, IJCAI, ASE, ACM MM, ICASSP.
\end{IEEEbiography}

\begin{IEEEbiography}[{\includegraphics[width=1in,height=1.25in,clip,keepaspectratio]{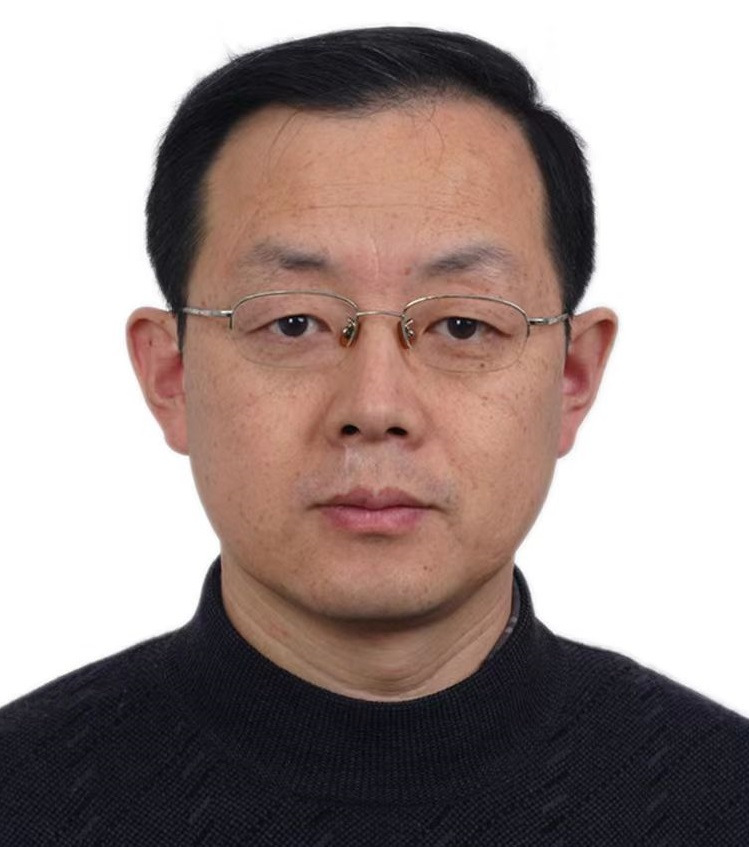}}]{Yong Dou}
Professor and Ph.D. supervisor with the National Key Laboratory of Parallel and Distributed Computing, National University of Defense Technology. His research interests cover high performance computing, intelligent computing, machine learning, and deep learning.
\end{IEEEbiography}

\vfill

\end{document}